 \documentclass[final,5p,times,authoryear]{elsarticle}

\usepackage{amssymb}
\usepackage{lipsum}
\usepackage{multirow}
\usepackage{url}
\usepackage{booktabs}

\journal{Astronomy $\&$ Computing}

\begin{document}

\begin{frontmatter}



\title{Experiences of Commercial Supercomputing in Radio Astronomy Data Processing}

\author[one,two]{I. P. Kemp\corref{cor1}}
\ead{ian.kemp@postgrad.curtin.edu.au}
\cortext[cor1]{Corresponding author.}
\affiliation[one]{organization={International Centre for Radio Astronomy Research (ICRAR)}, addressline={Curtin University, Bentley, 6102, WA, Australia}}

\affiliation[two]{organization={CSIRO Space and Astronomy}, addressline={26 Dick Perry Avenue, Kensington, 6151, WA, Australia}}

\author[one]{S. J. Tingay}

\author[six]{S. D. Midgely}
\affiliation[six]{organization={Defence Science and Technology Group}, addressline={Australian Department of Defence, Australia}}

\author[five]{D. A. Mitchell}
\affiliation[five]{organization={CSIRO Space and Astronomy},addressline={Cnr Vimiera \& Pembroke Roads, Marsfield, 2122, NSW, Australia}}

\begin{abstract}
The ongoing exponential growth of computational power, and the growth of the commercial High Performance Computing (HPC) industry, has led to a point where ten commercial systems currently exceed the performance of the highest-used HPC system in radio astronomy in Australia, and one of these exceeds the expected requirements of the Square Kilometre Array (SKA) Science Data Processors.

In order to explore implications of this emerging change in the HPC landscape for radio astronomy,  we report results from a survey conducted via semi-structured interviews with 14 Australian scientists and providers with experience of commercial HPC in astronomy and similar data intensive fields.  We supplement these data with learnings from two earlier studies in which we investigated the application of commercial HPC to radio astronomy data processing, using cases with very different data and processing considerations. 

We use the established qualitative research approach of thematic analysis to extract key messages from our interviews. We find that commercial HPC can provide major advantages in accessibility and availability, and may contribute to increasing researchers' career productivity. Significant barriers exist, however, including the need for access to increased expertise in systems programming and parallelisation, and a need for recognition in research funding. We comment on potential solutions to these issues.

\end{abstract}




\end{frontmatter}




\section{Introduction} \label{Introduction}

\subsection{Purpose of this study} \label{Purpose}
This work is part of ongoing research to explore the applicability of commercial supercomputing to next generation radio astronomy. In this paper we will show that multiple commercial computing providers could today host infrastructure with processing power of the scale required for significant future applications, including the Square Kilometre Array (SKA) Science Data Processors (SDP). We argue that raw processing power is not sufficient, and that there are a range of advantages and disadvantages associated with commercial High Performance Computing (HPC) which would need to be considered if this source of computing power is to find a role in radio astronomy.

This \textit{note on practice} has an Australian focus, on the grounds that Australia is currently on the cutting edge of radio astronomy instrumentation, with the Australian Square Kilometre Pathfinder (ASKAP), the Murchison Widefield Array (MWA) and the forthcoming low frequency telescope of the Square Kilometre Array (SKA-low). The Australian research community has extensive experience at the intersection of HPC and large-scale radio astronomy, and has been involved in planning future development in this area via the Australian SKA Regional Data Centre (AusSRC) survey \citep{AusSRC_Survey} and the AAS Decadal planning process \citep{AAS}. In this regard the Australian radio astronomy community may not be unique but its experiences may be representative, and help inform the wider international community as we adopt a new generation of radio telescopes.

To identify the key advantages and issues of commercial HPC, we report results from a qualitative research study carried out via interviews with a number of Australian researchers with recent practical experience of commercial supercomputing in Australia. We have not intended to develop an authoritative summary of the views of the Australian radio astronomy community. Also, we have not sought to collect data on or critique existing widely-used publicly provided HPC systems, nor provide a comparison between public and commercial HPC. We hope to disseminate to the wider radio astronomy community some of the experiences and lessons learned by researchers with practical experience of commercial HPC.

We have applied a rigorous qualitative research technique borrowed from the social sciences. We have generated knowledge by interviewing astronomers and computing specialists who have used commercial supercomputing in astronomy or similar data-intensive science, to capture their learnings and reflections. We have also incorporated learnings from two of our own recent case studies using commercial HPC.

\subsection{Real-time data processing}
Modern radio astronomy is critically dependent on HPC. As an example, the 36-dish Australian Square Kilometre Array Pathfinder (ASKAP) in Western Australia \citep{2012SPIE.8444E..2AS, 2021PASA...38....9H}, sends up to 2.5 GB/s of visibility data to a Science Data Processor system (SDP), which carries out calibration, imaging, source finding and cataloguing, and stores the processed data to the CASDA repository \citep{2020ASPC..522..263H}.  All data are processed through the SDP, which is implemented on the Setonix supercomputer at the Pawsey Supercomputing Centre in Perth\footnote{\url{https://pawsey.org.au/}}. This machine, rated at 27 PFLOP/s processing power, is the most performant publicly disclosed computer in Australia, and at November 2024 appeared as number 45 in the world Top500 list \citep{TOP500}. It is a general purpose system used for a variety of other processing.  The ASKAP SDP has access to 180 dual 2.45GHz AMD EPYC ``Milan'' 64-core CPU nodes, each delivering 2.5TFLOP/s \citep{Trader2021}, so we estimate that the SDP requirement is met with 450 TFLOP/s.

The second major SKA precursor in Australia, the 256-tile Murchison Widefield Array (MWA) (\citet{2009IEEEP..97.1497L}, \citet{2013PASA...30....7T}, \citet{2018PASA...35...33W}), is capable, with its recently upgraded correlator, of sustainably sending 22.8GB/s \citep{MWAX}, however in its normal operating modes it transmits 2.3GB/s. In contrast to ASKAP, visibility data from MWA is sent direct to archive at the Pawsey centre. Basic processing including calibration and RFI flagging is done later, either using the All-sky Virtual Observatory (ASVO) system implemented on Setonix\footnote{\url{https://asvo.mwatelescope.org/}}, or by astronomers using their own systems (which may include users' merit based time allocation on Setonix itself). The MWA ASVO system processes data via 10 CPU nodes, which we estimate provide 25 TFLOP/s. Note that the ASKAP SDP needs to be able to process data in real time, whereas the MWA equivalent processes data on demand.

Data flows and SDP processing requirements are soon to increase by two orders of magnitude as the next generation of radio telescopes are implemented. The SDP for the Next Generation Very Large Array (ngVLA, currently at the design stage) will be required to process visibility data at up to 133 GB/s \citep{ngVLA-Req}. The 197-dish Square Kilometre Array Mid-Frequency telescope (SKA-Mid, currently under construction) will send up to 292 GB/s of visibility data to the SDP \citep{2022SPIE12182E..0QM}, which will implement an expanded set of workflows to produce a wider range of science-ready data products than does today's ASKAP \citep{2014era..conf20201N}.

The SKAO organisation currently estimates that the SDP for each of the two SKA Phase I telescopes would be appropriately sized at 135PFLOP/s \citep{SKAO}.

\subsection{Non-real-time data processing}
The processing of data streaming from the telescope is only one use of HPC. Astronomers carry out extensive processing to extract scientifically useful information from the standardised science ready products generated by the SDP, or to carry out custom calibration and imaging. This processing presents a wide and unpredictable range of requirements.

Some useful data on non-real-time processing were provided by the AusSRC Radio Astronomy Data User Community Survey Report \citep{AusSRC_Survey}. Below we re-plot two charts summarising aspects of the survey responses of 74 members of the Australian radio astronomy community.

\begin{figure}[hbt!]
\centering
\includegraphics[width=\linewidth]{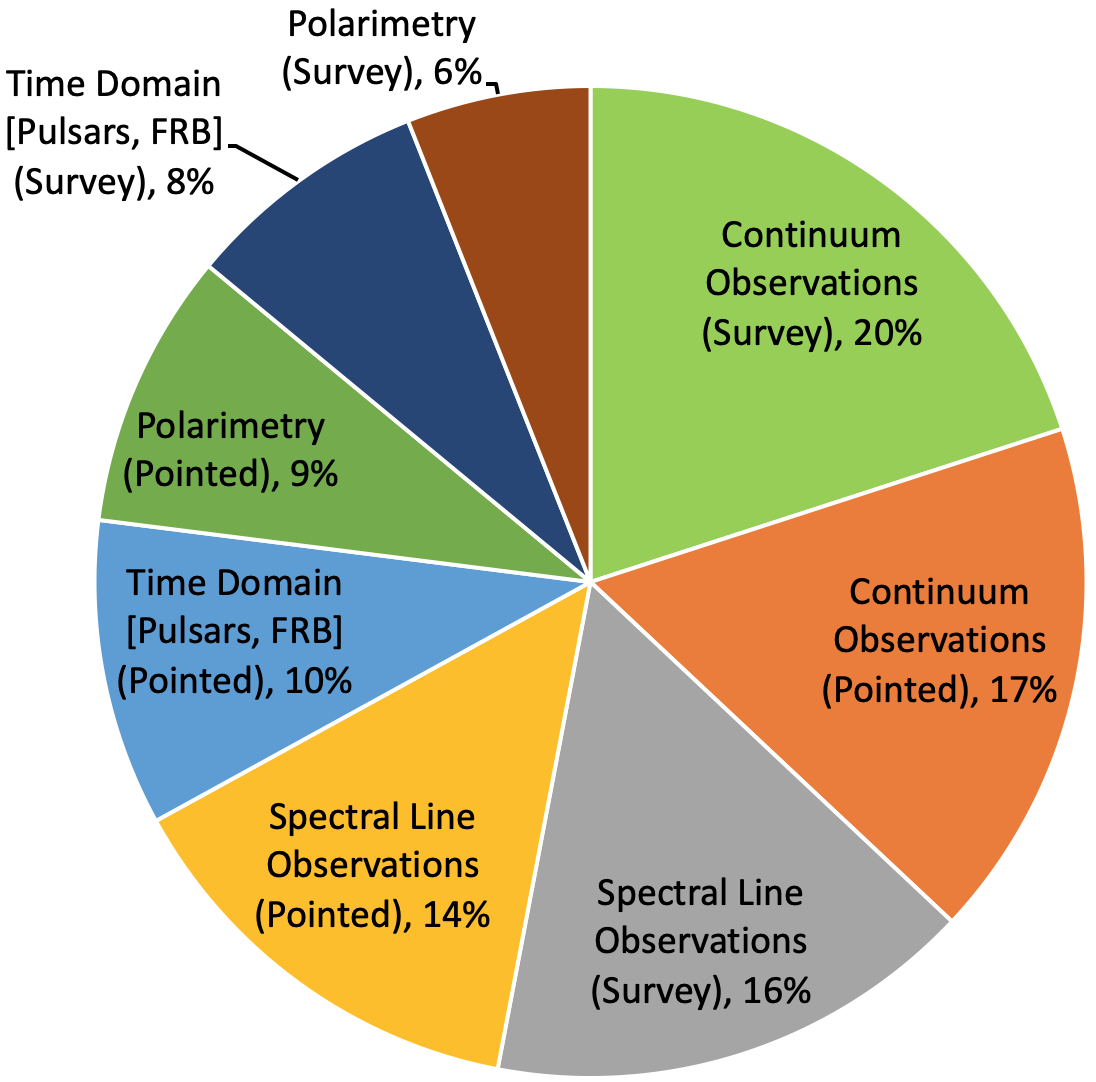}
\caption{``What is the observational nature of the data you are using?" Replot of Fig. 6.2.4 from the Australian SKA Regional Centre (AusSRC) survey \citep{AusSRC_Survey}.}
\label{fig_AusSRC-UseCase}
\end{figure}

\begin{figure}[hbt!]
\centering
\includegraphics[width=\linewidth]{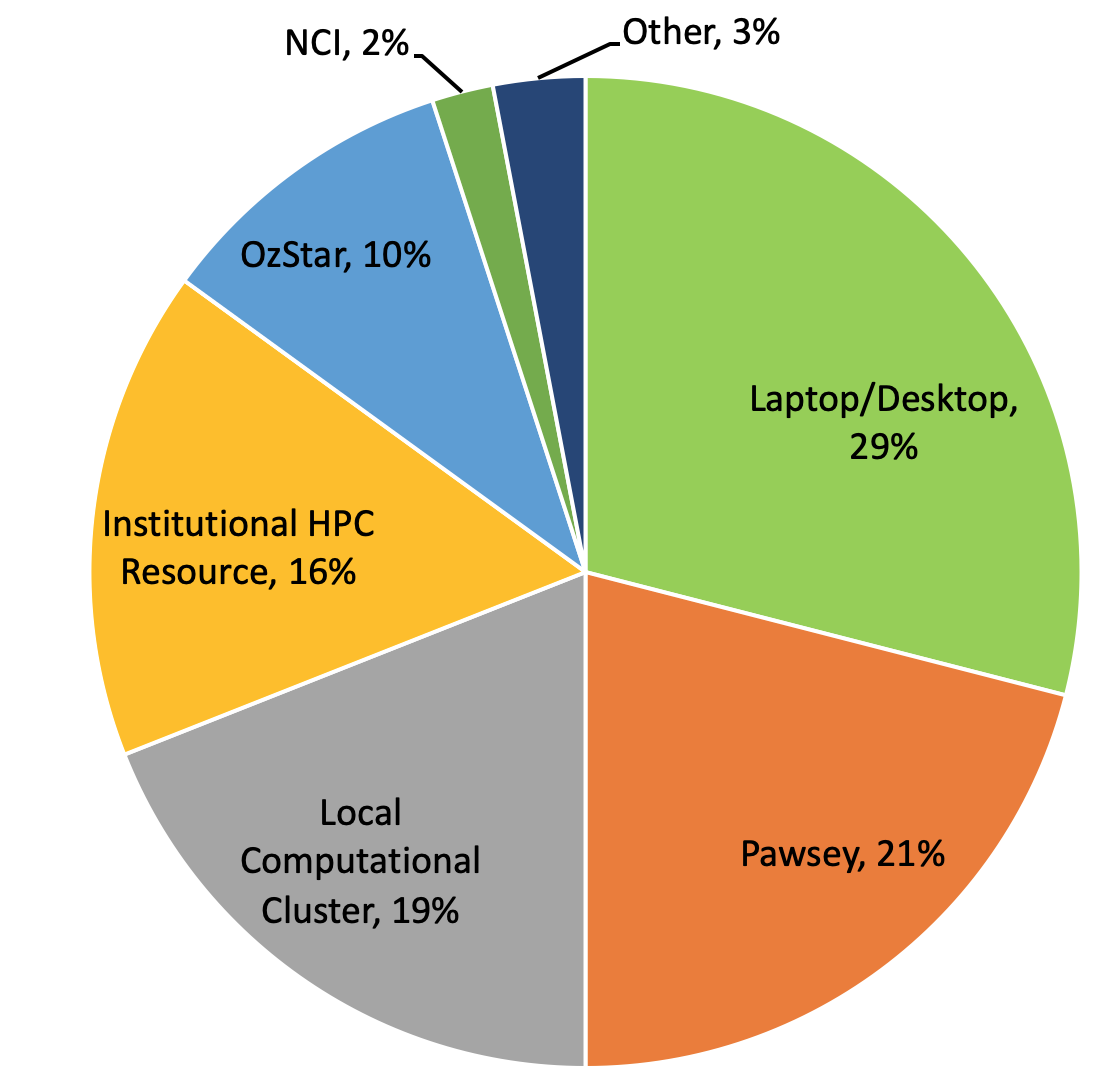}
\caption{``Where do you process your data?" Replot of Fig. 6.5.1 from the AusSRC survey \citep{AusSRC_Survey}.}
\label{fig_AusSRC-HPC}
\end{figure}

Figure \ref{fig_AusSRC-UseCase} illustrates the range of use cases for processing data, with differing technical requirements. For example, `pointed spectral line observations' using ASKAP may involve the processing of a small number of large data files. Processing H\textsc{i} images from the Galactic-ASKAP (GASKAP) pilot survey involved processing 6.6 TB data items, calling for nodes with large amounts of configurable RAM  \citep{2025A&C....5100901K}. Processing of time domain survey data may require the production of large numbers of snapshot images, favouring massively parallel processing using large numbers of multi-core nodes \citep{2024AJ....168..153K}. A search for specific object types in continuum survey data may require large numbers of transactions with archived data repositories \citep{2023Natur.619..487H}.

Most non-real-time processing (more than 97\% of responses) is done using government provided or institution-provided facilities, as shown by Figure \ref{fig_AusSRC-HPC}. Therefore publicly-provided HPC platforms are called upon to service a wide range of use cases with differing user requirements. 

Looking forward to the SKA, the need to service this wide range of requirements, together with the large processing power required for the SDP, has led the project to propose separation of the real-time and non-real-time requirements in the technical design. The current plan is for the dedicated SDP to processes incoming data to a range of pre-agreed science products \citep{2014era..conf20201N}, which will then be distributed to a network of SKA Regional Centres (SRC) for non-real-time data discovery, data processing, and visualisation \citep{2024ASPC..535..399S}. Currently SRCs are being planned for Canada, Europe, China, India, and Australia \citep{SKAOSRC}, with the scope and capabilities established to meet the needs and interests of the regional research communities.

The split of work between the SDP and SRC network is not yet finalised, nor the computing requirement or the technical or commercial approach to provisioning it. Decisions on these factors will be made in the 5-10 year time frame, and for the Australian AusSRC \citep{AusSRC_DSP}, the way forward will be influenced by future forms of the National Research Infrastructure Roadmap \citep{NRIR} (updated every five years), and the National Digital Research Infrastructure Strategy \citep{NDRIS} (published in 2024 with a 5-10 year scope).

\subsection{HPC industry trends} \label{SKA_Data_Challenge}
\begin{figure}
\centering
\includegraphics[width=\linewidth]{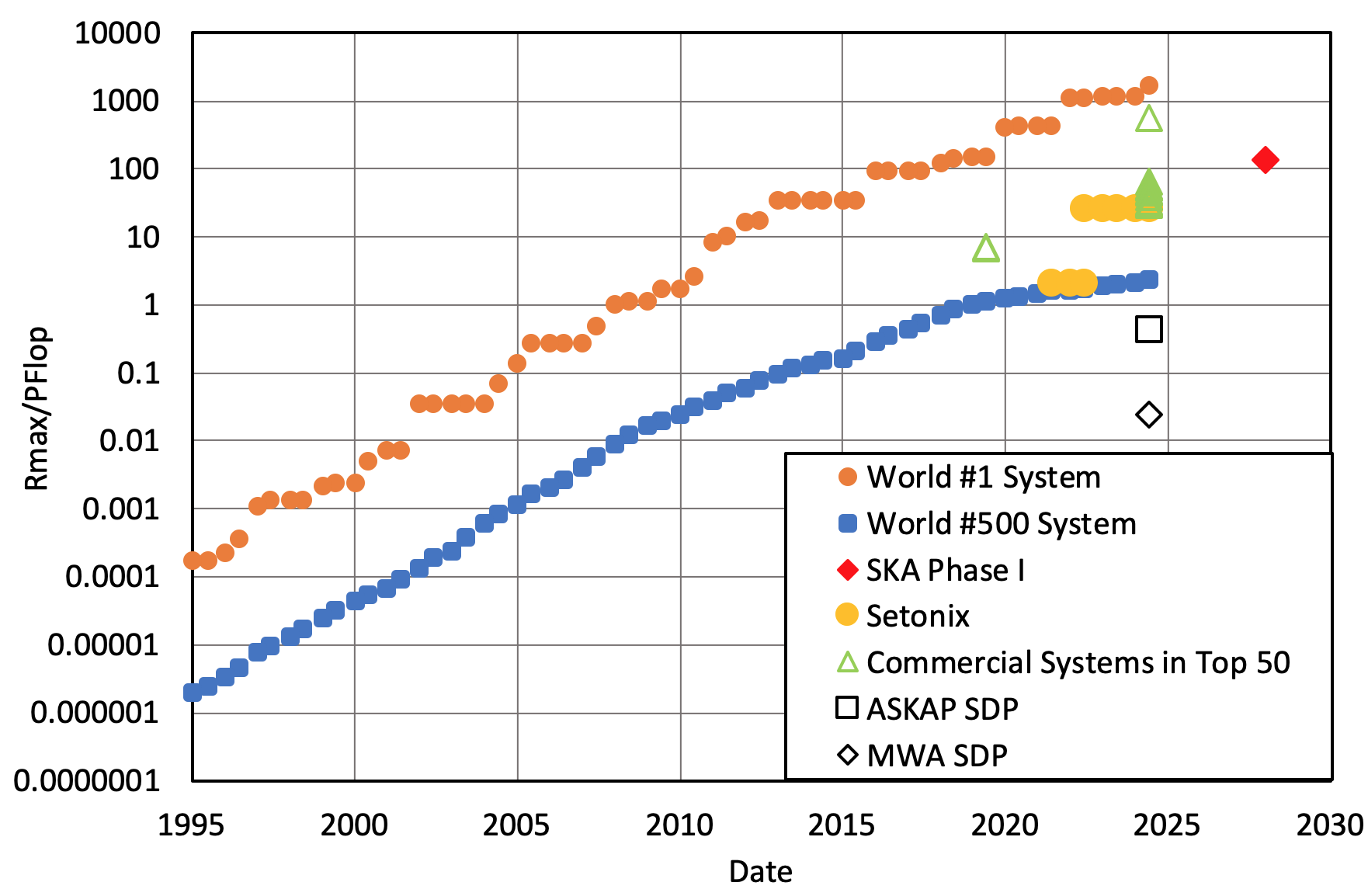}
\caption{Progression of the World \#1 and \#500 most performant computers \citep{TOP500}; Commercial systems in Top 50 in 2019 and 2024 \citep{TOP500}; and today's estimates of  SDP processing power for ASKAP, MWA (our estimates), and SKA-mid \citep{SKAO}.}
\label{fig_SKA-Top500}
\end{figure}

Early concerns that the performance requirement for the SKA SDP was beyond the capability of the world's largest supercomputers \citep{Norris2011} have been mitigated by the ongoing exponential increase in computing power, and extensions to the construction date of the SKA. In Figure \ref{fig_SKA-Top500} we plot the performance metric $R_{max}$ of the top system and the 500th system listed by Top500, and indicate the current estimate for the Phase I SKA-Mid SDP.  We acknowledge that raw processing power is only one aspect of performance, especially for data-intensive processing as required in radio astronomy, but here we use the raw performance (PFlop/s) to rank systems, recognising that this may not wholly correspond to performance in real-world applications.

Examination of the time-series Top500 survey data \citep{TOP500} reveals an increasing presence of large-scale commercial supercomputing systems.  Five years ago, in November 2019, two systems providing commercial cloud computing appeared in the top 50, the largest being the Artemis system in the United Arab Emirates, with an $R_{max}$ metric of 7.3 PFlop/s. In November 2024, ten commercial systems appeared in the top 50, the largest being the Eagle system provisioning Microsoft Azure, with a performance of 561 PFlop/s - placing it at the \#4 machine on the Top500 list. All of these ten systems have a higher performance rating than Setonix, the primary system used for radio astronomy data processing in Australia (Figure \ref{fig_AusSRC-HPC}), and Eagle exceeds the published requirement for the SKA SDP.

The performance of commercial systems in the top 50 of the Top500 list are also plotted in Figure \ref{fig_SKA-Top500}, along with our estimates of the processing power allocated to the ASKAP and MWA SDPs. This plot shows how top-end commercial HPC has overtaken the present-day requirements of large scale radio astronomy processing.

\subsection{Potential role of commercial HPC}

As reviewed above, it is now the case that multiple commercial computing providers could host platforms of the scale required by both real time and non-real-time processing in radio astronomy.

The advent of large modular systems built from standardised components (for example the Nvidia SuperPOD \citep{TechDay}) may make increasingly powerful platforms available to commercial providers in the future, at potentially lower cost.

Therefore, it is possible for commercial supercomputing to play a role, as a source of additional capacity either as a top-up or as an alternative to publicly-owned systems. Due to ongoing growth in the industry, this is likely to continue into the lifetime of the next generation radio telescopes currently planned or under construction. 

However, as reviewed above, radio astronomy presents a range of use cases and the applicability of commercial supercomputing will depend on factors beyond the simple $R_{max}$ performance metric. In this study we explore in more detail the key advantages and disadvantages which will determine the application of commercial supercomputing to radio astronomy. 

\subsection{Contributing studies} \label{contributing}
Recently the current authors participated in two technical studies, a search for fast radio bursts (FRB) using the Murchison Widefield Array (MWA) \citep{2024AJ....168..153K}, and a study of the dynamics of H\textsc{i} gas in our local galaxy group, using the Australian Square Kilometre Array Pathfinder (ASKAP) \citep{2025A&C....5100901K}. These two studies represented technically different types of research, and were carried out using the commercial computing resources of DUG Technology in Perth\footnote{\url{https://dug.com/about-dug/}}.

In addition to the science outcomes, these studies provided useful experience with commercial HPC, which helped guide and frame the questions for the survey presented in this paper. In the FRB study, a key success factor was the ability to optimise our search code to match the machine architecture.  In the second study, we formally extracted and published lessons learned, of which the key lessons were that the benefits of high accessibility and availability could be obtained only by accessing a higher than normal degree of computing skills. Please refer to the full publication for a more granular analysis of benefits and costs \citep{2025A&C....5100901K}.

\subsection{Structure of this paper} \label{Structure}
The structure of this paper is as follows.
In section \ref{Data_And_Methods} we explain our interview-based method of data collection, and our sources of information. In section \ref{Results} we lay out the themes extracted from the interviews and illustrate them with a mixture of paraphrase and direct quotation. In Section \ref{Discussion} we review the themes and extract learnings which are summarised in Section \ref{Conclusion}.

\subsection{Note on Terminology}
In this work we refer to `Bare Metal' and `Cloud' HPC. In `Bare Metal' we mean a service where the user pays for exclusive access to physical processors with only the operating system installed. By `Cloud' we refer to a virtualised service where virtual servers are assembled based on the user's specification.

We use the terms `real time' and `non-real-time' to describe different types of data processing. In this paper `real time processing' is referring to the primary processing of data streamed from a radio telescope, which is a necessary part of telescope operations. The term includes batch processing not strictly carried out in real time. `Non-real-time' indicates processing without a need to keep up with the data stream from the telescope, and is the normal mode for science research using archived data.

\section{Data and Methods} \label{Data_And_Methods} 

\subsection{Thematic Analysis} \label{Thematic_Analysis}
We use a qualitative research method to capture learnings from, and attitudes to, the use of commercial supercomputing in astronomy in Australia. Qualitative methods are commonly used in modern social science to extract meaningful information from survey data, where the information of interest can not readily be summarised numerically or graphically \citep{Merriam2015}.

Our primary tool is Thematic Analysis (TA), a technique common in qualitative social science research \citep{Clarke2012}. This methodology was originally formalised for psychological research \citep{2006Braun}, and to emphasise its importance in the qualitative research domain we note that Braun \& Clarke's 2006 paper was recently identifed as the 3rd most-cited paper of the 21st century \citep{2025Pearson}.

We use TA to study transcripts from semi-structured interviews, with open-ended questions, to identify common themes in the responses. This is not an exercise in word counting, or grouping together similar responses. Rather, the aim is to identify the (often unstated) issues of concern to the interviewees which underlie their responses to the questions. The power of this technique depends primarily on how common a theme is in the population of respondents, and the number of interview subjects.

Using TA, a strongly prevalent theme may be identified with a surprisingly small number of respondents. The relationship between prevalence, sample size and number of instances (number of times the theme is noted) was investigated in detail by \cite{2015Fugard}. As an example, these authors' Table 1 shows that in a sample of just 10 interviews, a theme with more than 50\% prevalence in the population will be detected (with 80\% confidence) by appearing in results from four or more interviews.

\subsection{Participants}
The focus of our study was radio astronomy, but to broaden the perspective and range of experience we included some researchers working in non-astronomy data intensive research, who had prior exposure either to radio astronomy or to the same computing platforms used by the astronomers. The participants had a range of experience of commercial HPC, and with differing levels of research experience and HPC proficiency.

We also accessed some experience from the field of geophysics, which is relevant because processing of geophysical data (seismic data) into maps and models is a similar process to that used in radio astronomy, and involves stacking of realtime data from multiple sensors, and multiple cycles of calibration and noise attenuation, to produce image cubes which can be explored in multiple ways to extract worthwhile information about the subsurface \citep{Saleh2011}.

We interviewed a total of 14 participants, as follows:
\begin{itemize}
\item{Four researchers in astronomy, and one professional staff member, from the Curtin University node of the International Centre for Radio Astronomy Research;}
\item{One researcher in astronomy, from the Australian National University (ANU);}
\item{One researcher in astronomy, from the Commonwealth Scientific and Industrial Research Organisation (CSIRO);}
\item{Two researchers in climate science, from CSIRO;}
\item{Two staff members from DUG Technology, a bare-metal HPC service provider in Perth;}
\item{Two staff members from a large international cloud computing provider doing business in Australia;}
\item{One former manager from the oil industry with experience of bare-metal HPC.}
\end{itemize}

Some of the participants requested anonymity, although the majority did not. To preserve anonymity of those requesting it, and to encourage candid responses, we use single-letter identifiers as listed in Table \ref{table:participants}, and have withheld names, job titles, and the identity of the cloud provider. The Table includes research experience and HPC experience, elicited in the interviews as described in Subsection \ref{Interviews}. The three non-corresponding co-authors of this paper were included as interview subjects, and so experience from our two case studies referred to above are brought in via interview responses.

Researchers and professional staff at ICRAR/Curtin were recruited via a call for participation at a regular all-staff meeting.  The CSIRO and ANU staff had taken part in a CSIRO trial of commercial supercomputing with DUG Technology, during 2022-2023. All CSIRO staff involved in the pilot were contacted by email, and we interviewed all those who self-selected.  We directly approached the HPC specialists from DUG Technology, and from the cloud computing provider. These staff from the providers were known to the research team having been involved in collaborative work with Curtin University.

Participants had experience, between them, of using Bare Metal and Cloud High Performance Computing. All the university and provider staff had obtained experience relevant to the interview in Australia. The oil industry representative had experience with bare-metal HPC in Australia and the United States of America (USA), and had been involved in a major outsourcing initiative with a large scale cloud computing provider in the USA. This experience is relevant because of a similarity between the computing needs and processes in geophysics and radio astronomy. 

Because of the way the participants were selected, most would be known to each other, however for this study the identities of participants were not disclosed or discussed with other participants.  Six of the 14 participants are based in Western Australia (WA), however, we have not investigated whether there is a systematic difference between attitudes in the WA radio astronomy and HPC community, and the wider community in Australia or globally. Therefore there remains the possibility of our results being biased by the relatively large WA participation.

It will be noted that our study did not include any subjects who have \textit{not} used commercial supercomputing. Therefore from this work we are not able to comment comparitively on the motivations for using or not using commercial providers.

\begin{table*}[hbt!]
\centering
\begin{tabular}{cllcc}
\toprule
Identifier & Role & Organisation & Research Experience & HPC Experience\\
\midrule
A & PhD Student, Astronomy & University & Low & High\\
B & Researcher, Astronomy & University & Low & Low\\
C & Researcher, Astronomy & University & High & High\\
D & Researcher, Astronomy & University & High & High\\
E & Researcher, Astronomy & University & High & Low\\
F & Professional, Astronomy & University & Low & High\\
G & Researcher, Astronomy & CSIRO & High & High\\
H & Researcher. Climate Science & CSIRO & High & High\\
I & Researcher, Climate Science & CSIRO & High & High\\
J & Manager & Bare Metal HPC provider & High & High\\
K & Professional & Bare Metal HPC provider & Low & High\\
L & Professional & Cloud provider & Low & High\\
M & Professional & Cloud provider & Low & High\\
N & Manager & Oil Industry & High & Low\\
\bottomrule
\end{tabular}
\caption{Summary of survey participants}
\label{table:participants}
\end{table*}

\subsection{Interviews} \label{Interviews}
The open-ended questions used in the interviews are listed in \ref{Survey_Questions}.  For each interview the question set was chosen to match the role of the respondent (Researcher, Manager, HPC provider).

Early questions were intended to elicit the degree of research experience and computing knowledge, which are indicated in Table \ref{table:participants}. In our coding, `High' research experience indicates more than three years postdoctoral research in science; `High' HPC experience indicates that the participant disclosed they themselves had set up the computing environment and/or significantly altered the software or job scripts to maximise performance on their chosen platform.

Later questions were intended to prompt reflection on the participants' experience and to elicit thoughts on the possible role of commercial supercomputing in the future.

In all cases but one, interviews were carried out online using Microsoft Teams video conferencing software, the single exception being one face to face interview.  Audio recordings were transcribed using Rev\footnote{https://www.rev.com/}, manually error corrected, and summarised manually into a digest. The purpose of the digest was to remove duplication and digressions, and thereby provide rapid access to relevant parts of the full transcript. Subjects were given the opportunity to correct the transcript and digest; corrections at this stage were very minor.

Transcripts were analysed using the Thematic Analysis methodology as described by \cite{Clarke2012}.  Initially, codes were developed using the transcript digest. Codes are one or two word terms developed by trial and error to label the topic being discussed at each point in the interview, for example `Funding', `Project Management', and `Skills'. Good granularity was obtained using 17 codes (listed in \ref{Codes}). 379 specific quotes were extracted from the original transcripts, each pertaining to one or more codes.

Themes were identified by filtering quotes on the codes, and examining these filtered lists in conjunction with the participants' research and HPC experience. Commonalities in the responses are identified as themes, and the themes and example quotes supporting them are discussed in the next section. The co-authors who were interview subjects were not involved in coding and theme extraction, so the thematic analysis was independent of their responses.

Due to the open-ended nature of the survey questions, participants had wide latitude to raise matters. For example the question `what difficulties did you experience’ elicited replies relating to software, system configuration, project management, and budget. The coding methodology means that our results are discussed in terms of the common themes raised by the respondents, and not simply in terms of the questions posed in the interviews. Formally, analysis of the survey data was predominantly inductive, though of course at a higher level all responses were framed by the interview questions.

\subsection{Human Research Ethics} \label{Ethics}
Research was carried out within the the Australian National Health and Medical Research Council's framework for Human Research Ethics \citep{NHMRC}, and was approved by the Curtin University Human Research Ethics Office \footnote{Approval number: HRE2024-0182}.

\section{Results} \label{Results}
\subsection{Experience}
Nine of our 14 respondents had experience with DUG's bare metal HPC platform. Four had experience with cloud computing, and one person had used both in their research.

Researchers had used a variety of software packages common in the radio astronomy community, including MIRIAD \citep{1995ASPC...77..433S}, CASA \citep{2022PASP..134k4501C}, and WSClean \citep{2014MNRAS.444..606O}, as well as some more niche products. Where the researchers incorporated their own code, this was usually coded in Python using standard Python libraries. Two respondents reported using containerised packages. 

In some cases respondents detailed issues or problems in installing or running the software, but no common themes were identified and they did not appear to relate to the commercial character of the service.

\subsection{Themes}

In this section we describe the themes which were extracted from the coded interviews. We summarise each theme, provide detail including paraphrases and/or exact quotes from respondents, and further comment based on the current authors' own experience.

\subsubsection{Accessibility and Flexibility}
Respondents were unanimous in expressing an overall positive experience with commercial HPC. Most were able to identify technical problems, but noted that these were not specific to commercial HPC and would likely arise with any change to a new platform.  The main advantage of commercial provision, mentioned by six respondents, was flexibility - the ability of the users to obtain different processing nodes, larger memory and storage, on demand. F said: ``If you wanted to run something that needed 200 GB RAM ... they provisioned a high spec machine and you only needed it for that afternoon''. 

Five respondents mentioned availability - the commercial systems had very little `downtime': ``Whatever time of the day, whatever day of the week, I know I'm just going to be able to access'' [A]. The main value of high availability is in reducing the overall time for the research project: ``Private sector computers can be great because we ... just churn through the data processing so we can get to the science stage fastest'' [B].

Mentioned by three respondents, was the ability to use commercial HPC as overflow capacity, to supplement provision from public systems at certain times, although there is a cost in time and effort expected with setting up the code and data management on a new platform: ``There is overflow availability because places like Pawsey are always going to be over subscribed. Rapid availability can be a thing, there’s always an opportunity for commercial overflow when [they can] provide say 4000 machines really quick'' [K].

Four respondents who had used the bare metal provider remarked on the technical support. H said: ``A major `pro’ was direct access to the optimisation specialists'', although two respondents (each with a high level of HPC experience and skill) said they required no technical support.

Eight respondents raised issues specific to commercial HPC. Five mentioned concerns about security: ``One of the dangers of working with external providers is concerns over data access and data integrity, though this concern may be misplaced, they might have better security than our in-house management'' [N]. Three mentioned concerns that processing on cloud systems may lead to data crossing national boundaries, in conflict with national data policies: ``In my space there are rules around data and it is very cloud unfriendly'' [J], while three respondents reported difficulties in data management, either the need to transfer data from its main repository, or being hampered by quotas to restrict the cost of storage: ``All of the input data has to be fed into [a] data lake, because it’s just prohibitive to be doing data transfers'' [H].

\subsubsection{One size doesn't fit all}

We did not ask about experiences with public facilities, but seven of the respondents commented on problems with their existing platforms, contrasting their experience with commercial HPC. A large public facility may lack flexibility in hardware configuration; F said ``A problem with using Setonix is that it’s AMD GPU based, a lot of our code uses Nvidia specific libraries''. Changes occur at infrequent major upgrades: ``Commercial providers generally have the latest and greatest hardware and bringing online new hardware constantly, national facilities get upgraded in a monolithic way'' [E].

Another issue raised with public systems was the risk to integrity and performance from sharing the system between users of a wide range of skills and experience, particularly in the case of real-time processing. M said: ``You are at the mercy of everyone doing the right thing'', while G, speaking about the implementation of the SDP for a major telescope, said: ``I think we need a chunk of nodes that we can isolate from the rest of the system and reboot and install ... whatever we need to run the telescope''.

In total, six respondents discussed the difficulty of using a single general-purpose installation for both real time and non-real-time processing. A preference for different platforms for different use cases then prompted a discussion on the best role for public vs commercial systems. In relation to the SDP, G said ``the computer hangs off the bottom of the telescope to do the data reduction, if there’s only one and you can never afford to have competition there’s no point in commercialising it''.  For discretionary computing, said E, ``a national facility that's free for users on a merit based allocation is going to be, in practice, probably the primary provider of compute and storage for most researchers.''

\subsubsection{We need to access advanced computing skills}
Eleven respondents commented on the need for system admin, system programming, and parallelisation skills beyond that normally taught to graduates in physics, because of the critical role of HPC in radio astronomy as in other sciences. However, there was no consensus on whether those skills should be brought by the researchers, or provided by the HPC provider, or a third party.

Three out of the five users with cloud experience commented on the need for HPC skills above that normal for an astronomy researcher, while six out of the ten users with experience of the bare metal provider mentioned the assistance from the HPC team as an unexpected but major help.

Speaking in favour of upskilling researchers, in all fields not just astronomy, M said: ``I think of the computing as the universal scientific instrument right now ... [using it] in an efficient way requires training and expertise", but H asked ``is that a good use of researcher time, or do you really want to have someone who’s job it is to provision commercial compute resources?'' 

E suggested ``What we need is a new cohort of people that sit in the middle [with a] high level awareness of the astronomy and a reasonably high level awareness of the computing ... the AusSRC should be just about the first time this has been attempted in Australia.''

Looking to the future, responses slightly favoured the idea that HPC vendors should provide access to a skilled team (four responses), rather than relying on upskilling researchers (two), or a specialist team in the research community (two). A further three respondents were agnostic, saying that any of the three approaches could be used to provide a basis of HPC skills.

Five of the nine respondents experienced with the Bare Metal provider commented on the value of a team approach, in which researchers and HPC specialists work together with some overlap of skills. The team approach with regular meetings fosters overlap of knowledge which improves the team efficiency. The researchers would need to ``be willing to converse and have a good dialogue with ... the industry side, ... so that everyone can get up to speed as quickly as possible'' [C], because ``[if] you are doing research ... you're unable to write a specification and ... it's good to be able to sit down with people who have understanding, and work out what [to do]'' [G].

A team approach with members from the HPC provider may work well if the providers ``get a bit familiar with some astronomy software packages to be able to maybe instal them themselves and maintain them rather than expect astronomers to do it'' [D]. This would be easier for a team using HPC specialists from the research community, provided they had sufficient knowledge of the specific providers' infrastructure.

\subsubsection{Costs and funding aren't aligned}
Five respondents made the point that, with publicly-provided HPC, costs were hidden and the service presented as free to the users, whereas with a commercial provider, costs are more explicit and can spur improvements in efficiency.  M pointed out that this can lead to academic users exaggerating the cost of commercial HPC: ``Traditionally with academic circles there was a focus on hardware cost, because power, labour, security, data centre space, were all free. In a cloud provider those costs are all rolled in so people tend to compare just hardware cost with cost of everything.''

On the other hand, the need to pay for computing ``can be a pro, as it forces people to think about their usage; when something comes for free usage is pretty lazy and expedient'' [E].

There was a range of opinions on the actual costs of the commercial providers. F recommended close attention to the cost model and the actual services provided: ``They're very good at pushing costs down to the end user ... it's very easy to end up paying for things that you don't need''. Management of a service contract may be an unfamiliar skill for academics. ``Cost is definitely scary, especially for academics where we're always clamouring for money. ... But then if you look at other fields of science, ... they get their funding to build their lab or to get their lab equipment'' [B].

The actual service catalogue and cost model of the provider may not fit well with the users' needs in a specific project; for example ``pricing model based on ... monthly storage ... doesn’t lend itself to radio data processing, because you have so many intermediate data products'' [C].

These comments reflect a concern that a commercial provider may be motivated to over-charge, or charge for unnecessary services. But one of the providers said that in practice users tend to use all of the project budget regardless, so should ask for help to improve efficiency. ``In Astronomy, if you run someone’s code 10 times faster, they’ll probably give you the same amount of money because they’re just going to get 10 times as much imaging done. I’ve spoken to a lot of astronomers and they’ve never said `we ran out of data’ '' [K].

Five respondents mentioned that the current funding model for astronomy in Australia presents some barriers to the use of commercial supercomputing.  ``Our telescopes and our computers are funded separately, managed separately, operated separately, and have to ... coordinate to achieve the overall objectives ... In the future ... [funding] needs to be fundamentally attached to a project'' [E].

Respondents commented on the difficulty of funding long term work, and in justifying expenditure, both issues which are unlikely to be unique to HPC.  ``Cost is a bit prohibitive because for science projects you don’t have a guarantee that you’ll have this funding over time to cover the costs'' [D].  ``I’ll need to justify why we are using a more expensive machine, ... what value are we getting out of that'' [I].

\subsubsection{Software is becoming a big problem} \label{Software}
Ten respondents chose to engage in a discussion about software. Specific issues with common software all reflect that installing and optimising code on a new platform was a barrier to using alternate infrastructure or overflow capacity.  ``[WSClean is] a difficult software package to compile, especially on a cluster environment where libraries need to be shared among submission vs compute nodes'' [C]. ``One of the big problems ... was resolving all the complex dependencies for MIRIAD to get it compiled. I'm really lucky ... I don't have to run other people's code'' [A].

The effort to move software to a new platform counteracts some of the key potential advantage of commercial HPC, the ability to switch providers to get a better deal, and to rapidly provide overflow capacity. ``There’s all these models interacting, ... so when I ask someone `can you go in and optimise the code’ they look at it all and go `where do we even start!’ ... it works better if we’ve had a long term relationship [with the provider]'' [I].  ``The problem of HPC is to take application software and make it understand the underlying hardware and map onto the machine efficiently. ... Python ... doesn’t have the concept of parallelism. We need to retrain people in higher performing languages like Go and Rust'' [N].

Discussions on deployability reinforced the concept of having a dedicated team to manage a software base for the astronomy community, although funding is seen as a barrier. ``How do you fund it in the academic space? Keeping 20 people who are going to take all the stuff that people do and convert it into a proper commercial software product?'' [J].  ``You (could) pay some of your grant money for these people to give you your science products or your data products that you can do science with'' [G].
``I've never been able to figure out what a commercial model for it would actually look like. I suspect it would end up being essentially something like a software developer Patreon for astronomy software because any given project doesn't use all the functionality of the software [K].''

\subsubsection{We can learn from the field of geophysics}

Our respondent from the Oil and Gas industry brought some useful comments based on experience of outsourcing geophysical data processing to the private sector. Initially seen as risky from the point of view of data security and integrity, the company ended up following the value in having the data centre and software development run by (separate) private companies for whom it was their core business. The complexity of managing key software was handled by contributing code and methods to a third party company, which multiple oil companies support by licensing software. Thus key basic functionality and algorithms are maintained in a professional manner, while the oil companies keep proprietary control of some imaging and search methods which they consider a competitive advantage.

N said: ``[The providers'] core business is [high performance computing], so they will be developing and keeping abreast of new technology, developing their own tools, improving their own tools, their capability, and always their business case relies on them being able to deliver a cost-effective reliable and efficient service.'' The geophysicist believes that the outsourced model has the advantage that ``commercial providers are able to put a lot more people into developing code and meeting client’s needs. Researchers will develop their own tools for use in interpretation ... You know what you’re getting and it’s an industry standard that everyone can understand''.

Flexibility can be obtained even within a long term partner relationship: ''Contractors now are able to add nodes between computing centres around the world, like we’ve done stuff where ... we run it on their main computer in Paris that they have available time, or they’d run it somewhere else. This can be a problem if you don’t own the data, if it’s owned by a state or something. But generally these days you’re given the flexibility''.

It's important to emphasise that this model was implemented in the oil and gas industry which can apply significant funding in the pursuit of rapid and reliable data processing, and it may not be appropriate or even possible for the astronomy community.

\subsection{Notable comments} \label{sparklers}
Some notable points captured here were raised by individual respondents, but do not appear as themes in the joint analysis of the interviews.

In discussing cloud vs bare metal, respondent I commented that the relationship with the provider may be more important than technology: ``It might come down to that local relationship and the ability to address problems more quickly rather than a more impersonal operation ... you feel a bit more like being locked-in rather than being enabled, I guess''.

H commented: ``I have a moral objection to making money off research, and so I hate the idea that we should be putting aside money to pay commercial providers to do our work''. 

G highlighted that ``Nobel prizes associated with astronomy generally come from things that arrays weren’t initially built to do. With big public systems ... much of their time is already dedicated to big surveys, so if you want to do exploratory type processing ... the private clouds are another way to do that''.

In discussing the amount of resource required for an astronomy project, F made the valuable point that in reality ``[The] amount of computation that people do is always going to expand to fit the capacity that there is''.

In the following section we expand on the comments of our respondents and include our own specific experience from the two case studies referred to in Subection \ref{contributing}.

\section{Discussion} \label{Discussion}
We summarise and reflect upon our analysis in Section \ref{Introduction}, and interview responses in Section \ref{Results}, as follows:

Advances in computing technology and growth in commercial provision mean that it is feasible for the commercial sector to provide HPC for the most demanding cases in radio astronomy, including real time processing of data from a large array such as an SKA telescope.  Also the commercial sector can ‘retail’ HPC for smaller, non-real-time, applications and thus meet most potential needs of the radio astronomy community.

The experiences of our interview respondents reveal a number of specific advantages of commercial HPC, as compared to the traditional public systems at institutions or publicly run computing centres.

First among these is flexibility; with a range of providers each hosting a range of technologies, it is possible for the researcher to obtain a computing solution closely matched to their needs; for example CPU vs GPU, massive scalability, large or small node memory as required.  This is especially the case where the provider is able to constantly add the latest technology to their existing network, in contrast to a ‘monolithically’ upgraded public system.  The second major advantage is accessibility and availability: commercial providers appear to prioritise ease of access and very low downtime, which align with researcher’s contradictory needs to (a) get quick access to a large machine for an exploratory computation, and (b) obtain a stable platform for a large data processing task.

To access these advantages it is necessary to address barriers. Comments from interview participants and our own practical experience suggest that to take advantage of flexibility, the researcher must be able to evaluate and select among alternate computing platforms.

It is generally expressed in the commercial literature that bare metal HPC offers higher performance than cloud computing (for a given hardware platform), while cloud computing has a lower cost of entry (for example  \cite{Databank2025}, \cite{Servers2025}), and with cloud computing, a researcher may be faced with a choice of hundreds of possible virtual platforms. We reflect therefore that access to the skills required to make this type of selection may become a key factor in obtaining optimum value from a commercial provider.

Access to commercial HPC at scale requires payment, which may be difficult for academics to resource. Although commercial provision allows researchers to avoid the merit based time assignment process, there would instead be a need to obtain sufficient research funds by making the case for a commercial provider as an alternative to the existing public computing. Although HPC costs may not be large for a small project, resourcing these costs may be an unfamiliar process.

Another barrier is code portability, or, more accurately, access to skilled people able to manage a code base in the long term beyond its initial development, and who can provide installation and optimisation advice or a complete service. We have seen how a well-funded commercial sector (oil industry) has enabled this function to become a commercial industry in its own right. For research in radio astronomy this may not be possible, and as suggested by one of our interview subjects (section \ref{Software}) a small team funded via a `patreon model' may be workable (in which the team receives public funds indirectly via research projects).

There are also issues with data management - the use of commercial HPC which is not `close to' the observations can be a problem in processing greater than multi-TB of data. We remark that here is scope for HPC providers to assist with methods to transfer data more efficiently, and also to re-evaluate storage cost models, and for researchers to develop methods closer to `streaming' than `download and process'. There is some discussion of practical approaches in \cite{2025A&C....5100901K}.  Also, it may be difficult, especially in the case of cloud computing, to ensure that data are kept within the national boundaries which may be required by governments in return for funding hardware, operations, and data management.

To obtain the greatest benefit from the commercial sector, the community would need to seek changes, to allow some component of research funding to be applied to commercial HPC.  If the community were to decide to incorporate a large component of commercial HPC, our survey responses suggest this would probably best be done via a mixed model. In this, (a) funding for real time SDP-type processing would be an integral part of the telescope budget, and decisions on the model for provision would be part of the technical design options developed during the project; (b) for non-real-time processing, there could be a hybrid model, where a public resource (allocated as now via merit based time assignment) would be supplemented with a commitment to fund a certain amount of commercial HPC, via funds obtained through existing grant processes. Such a recognition of the commercial sector could enable the commercial providers to commit to long term support for the radio astronomy community as a specific customer group.

Finally we address the notable comments recorded in Section \ref{sparklers}. Researchers have their own individual perspectives and in some cases there appears to be a degree of suspicion about the motives and practices of commercial providers among some of our respondents.  However, `lock in' can be framed as `partnership' and we believe it is incumbent on the research community to avoid bias by seeking assistance if necessary to understand the capabilities of HPC providers and to select a provider best suited to their needs.

In response to the comment on the ethics of paying commercial providers, we note that at least in Australia, all public computing systems are ultimately supplied by the private sector, as is most non-computational research infrastructure - this is a reality of working in a mixed national economy.

The two comments on capacity reflect advantages of commercial HPC: systems can be provided quickly for small concept studies, while at the same time researchers can ride the curve of exponentially increasing HPC power to produce ever more sophisticated analysis of existing data. We note that, in combination, these factors may offer the prospect of reducing time-to-publication, and increasing the career productivity of researchers who's research output may currently be limited by the needs of data processing.

\section{Conclusion} \label{Conclusion}
In this paper we have used data on the continuing exponential development of HPC and the commercial sector in particular, to argue that commercial providers can provide infrastructure capable of performing radio astronomy data processing, even at large scale such as SDP processing for SKA telescopes.

Interviews with astronomers and other practitioners in data-intensive science revealed that commercial HPC offers a number of advantages for researchers in radio astronomy, including rapid provision, high availabilty, flexibility and continuous hardware upgrade. Barriers include a requirement for increased HPC skills, which need to be sourced, either by upskilling users, or sought from the HPC provider, or obtained via a third party such as AusSRC or ADACS \footnote{https://adacs.org.au/}. 

If the community would seek to use commercial HPC as a significant resource for the processing of radio astronomy data, approaches would need to be identified to some larger issues, including software maintenance, and a funding model which would permit easy access to commercial computing to supplement or replace publicly provided facilities. These aspects may be difficult for an academic community which resists standardisation and may be suspicious of the commercial sector, but potentially can provide significant benefits in terms of research output and lifetime productivity.

\section{Acknowledgements}
We thank our interview subjects for generously giving their time and expertise for this study.

\section{Funding}

This research was part of a CSIRO iPhD program, funded by CSIRO, Curtin University and DUG Technology as the industry partner.

\section{Data Availability}

In line with Human Research Ethics Guidelines, transcripts of interviews carried out for this research are confidential to the research team and, in the event of an audit or investigation, to staff from the Curtin University Office of Research and Development. Transcripts will be kept securely for seven years then destroyed.

\section{ORCID IDs}
\noindent
Ian Kemp https://orcid.org/0000-0002-6637-9987 \\
Steven Tingay https://orcid.org/0000-0002-8195-7562 \\
Daniel Mitchell https://orcid.org/0000-0002-1828-1969 \\

\bibliographystyle{elsarticle-harv} 
\bibliography{KempEtAl.bib}

\begin{thebibliography}{37}
\expandafter\ifx\csname natexlab\endcsname\relax\def\natexlab#1{#1}\fi
\providecommand{\url}[1]{\texttt{#1}}
\providecommand{\href}[2]{#2}
\providecommand{\path}[1]{#1}
\providecommand{\DOIprefix}{doi:}
\providecommand{\ArXivprefix}{arXiv:}
\providecommand{\URLprefix}{URL: }
\providecommand{\Pubmedprefix}{pmid:}
\providecommand{\doi}[1]{\href{http://dx.doi.org/#1}{\path{#1}}}
\providecommand{\Pubmed}[1]{\href{pmid:#1}{\path{#1}}}
\providecommand{\bibinfo}[2]{#2}
\ifx\xfnm\relax \def\xfnm[#1]{\unskip,\space#1}\fi
\bibitem[{{Australian Academy of Science}(2025)}]{AAS}
\bibinfo{author}{{Australian Academy of Science}}, \bibinfo{year}{2025}.
\newblock \bibinfo{title}{{Decadal plan for Australian astronomy 2026–2035}}.
\newblock \URLprefix \url{https://www.science.org.au/supporting-science/science-policy-and-analysis/decadal-plans-for-science/astro2035}. \bibinfo{note}{[Accessed: 6 Jun 2025]}.
\bibitem[{{Australian Government}(2021)}]{NRIR}
\bibinfo{author}{{Australian Government}}, \bibinfo{year}{2021}.
\newblock \bibinfo{title}{National Research Infrastructure Roadmap}.
\newblock \bibinfo{publisher}{Australian Government}, \bibinfo{address}{Canberra, Australia}.
\newblock \URLprefix \url{https://www. education.gov.au/national-research-infrastructure/ resources/2021-national-research-infrastructureroadmap}.
\bibitem[{{Braun} and {Clarke}(2006)}]{2006Braun}
\bibinfo{author}{{Braun}, V.}, \bibinfo{author}{{Clarke}, V.}, \bibinfo{year}{2006}.
\newblock \bibinfo{title}{{Using thematic analysis in psychology}}.
\newblock \bibinfo{journal}{Qualitative Research in Psychology} \bibinfo{volume}{3}, \bibinfo{pages}{77--101}.
\newblock \URLprefix \url{https://www.tandfonline.com/doi/abs/10.1191/1478088706qp063oa}.
\bibitem[{{CASA Team} et~al.(2022){CASA Team}, {Bean}, {Bhatnagar}, {Castro}, {Donovan Meyer}, {Emonts}, {Garcia}, {Garwood}, {Golap}, {Gonzalez Villalba}, {Harris}, {Hayashi}, {Hoskins}, {Hsieh}, {Jagannathan}, {Kawasaki}, {Keimpema}, {Kettenis}, {Lopez}, {Marvil}, {Masters}, {McNichols}, {Mehringer}, {Miel}, {Moellenbrock}, {Montesino}, {Nakazato}, {Ott}, {Petry}, {Pokorny}, {Raba}, {Rau}, {Schiebel}, {Schweighart}, {Sekhar}, {Shimada}, {Small}, {Steeb}, {Sugimoto}, {Suoranta}, {Tsutsumi}, {van Bemmel}, {Verkouter}, {Wells}, {Xiong}, {Szomoru}, {Griffith}, {Glendenning} and {Kern}}]{2022PASP..134k4501C}
\bibinfo{author}{{CASA Team}}, \bibinfo{author}{{Bean}, B.}, \bibinfo{author}{{Bhatnagar}, S.}, \bibinfo{author}{{Castro}, S.}, \bibinfo{author}{{Donovan Meyer}, J.}, \bibinfo{author}{{Emonts}, B.}, \bibinfo{author}{{Garcia}, E.}, \bibinfo{author}{{Garwood}, R.}, \bibinfo{author}{{Golap}, K.}, \bibinfo{author}{{Gonzalez Villalba}, J.}, \bibinfo{author}{{Harris}, P.}, \bibinfo{author}{{Hayashi}, Y.}, \bibinfo{author}{{Hoskins}, J.}, \bibinfo{author}{{Hsieh}, M.}, \bibinfo{author}{{Jagannathan}, P.}, \bibinfo{author}{{Kawasaki}, W.}, \bibinfo{author}{{Keimpema}, A.}, \bibinfo{author}{{Kettenis}, M.}, \bibinfo{author}{{Lopez}, J.}, \bibinfo{author}{{Marvil}, J.}, \bibinfo{author}{{Masters}, J.}, \bibinfo{author}{{McNichols}, A.}, \bibinfo{author}{{Mehringer}, D.}, \bibinfo{author}{{Miel}, R.}, \bibinfo{author}{{Moellenbrock}, G.}, \bibinfo{author}{{Montesino}, F.}, \bibinfo{author}{{Nakazato}, T.}, \bibinfo{author}{{Ott}, J.}, \bibinfo{author}{{Petry}, D.}, \bibinfo{author}{{Pokorny}, M.},
  \bibinfo{author}{{Raba}, R.}, \bibinfo{author}{{Rau}, U.}, \bibinfo{author}{{Schiebel}, D.}, \bibinfo{author}{{Schweighart}, N.}, \bibinfo{author}{{Sekhar}, S.}, \bibinfo{author}{{Shimada}, K.}, \bibinfo{author}{{Small}, D.}, \bibinfo{author}{{Steeb}, J.W.}, \bibinfo{author}{{Sugimoto}, K.}, \bibinfo{author}{{Suoranta}, V.}, \bibinfo{author}{{Tsutsumi}, T.}, \bibinfo{author}{{van Bemmel}, I.M.}, \bibinfo{author}{{Verkouter}, M.}, \bibinfo{author}{{Wells}, A.}, \bibinfo{author}{{Xiong}, W.}, \bibinfo{author}{{Szomoru}, A.}, \bibinfo{author}{{Griffith}, M.}, \bibinfo{author}{{Glendenning}, B.}, \bibinfo{author}{{Kern}, J.}, \bibinfo{year}{2022}.
\newblock \bibinfo{title}{{CASA, the Common Astronomy Software Applications for Radio Astronomy}}.
\newblock \bibinfo{journal}{PASP} \bibinfo{volume}{134}, \bibinfo{pages}{114501}.
\newblock \DOIprefix\doi{10.1088/1538-3873/ac9642}, \href{http://arxiv.org/abs/2210.02276}{{\tt arXiv:2210.02276}}.
\bibitem[{Clarke and Braun(2012)}]{Clarke2012}
\bibinfo{author}{Clarke, V.}, \bibinfo{author}{Braun, V.}, \bibinfo{year}{2012}.
\newblock \bibinfo{title}{Thematic Analysis}. \bibinfo{publisher}{American Psychological Association}, \bibinfo{address}{Washington, DC}.
\newblock pp. \bibinfo{pages}{57--71}.
\newblock \URLprefix \url{https://doi.org/ 10.1037/13620-004}, \DOIprefix\doi{10.1037/13620-004}.
\bibitem[{{Databank}(2024)}]{Databank2025}
\bibinfo{author}{{Databank}}, \bibinfo{year}{2024}.
\newblock \bibinfo{title}{High-performance computing: Cloud vs. bare metal}.
\newblock \URLprefix \url{https://www.databank.com/resources/blogs/high-performance-computing-cloud-vs-bare-metal/}. \bibinfo{note}{[Accessed: 31 May 2025]}.
\bibitem[{{Fugard} and {Potts}(2015)}]{2015Fugard}
\bibinfo{author}{{Fugard}, A.J.B.}, \bibinfo{author}{{Potts}, H.W.W.}, \bibinfo{year}{2015}.
\newblock \bibinfo{title}{{Supporting thinking on sample sizes for thematic analyses: a quantitative tool}}.
\newblock \bibinfo{journal}{Internation Journal of Social Research Methodology} \bibinfo{volume}{18}, \bibinfo{pages}{669--684}.
\newblock \URLprefix \url{https://doi.org/10.1080/13645579.2015.1005453}, \DOIprefix\doi{10.1080/13645579.2015.1005453}.
\bibitem[{{Hiliart}(2019)}]{ngVLA-Req}
\bibinfo{author}{{Hiliart}, R.}, \bibinfo{year}{2019}.
\newblock \bibinfo{title}{{{ngVLA} Computing and Software: Preliminary Requirements}}.
\newblock \URLprefix \url{https://ngvla.nrao.edu/download/MediaFile/ 243/original}.
\bibitem[{{Hotan} et~al.(2021){Hotan}, {Bunton}, {Chippendale}, {Whiting}, {Tuthill}, {Moss}, {McConnell}, {Amy}, {Huynh}, {Allison}, {Anderson}, {Bannister}, {Bastholm}, {Beresford}, {Bock}, {Bolton}, {Chapman}, {Chow}, {Collier}, {Cooray}, {Cornwell}, {Diamond}, {Edwards}, {Feain}, {Franzen}, {George}, {Gupta}, {Hampson}, {Harvey-Smith}, {Hayman}, {Heywood}, {Jacka}, {Jackson}, {Jackson}, {Jeganathan}, {Johnston}, {Kesteven}, {Kleiner}, {Koribalski}, {Lee-Waddell}, {Lenc}, {Lensson}, {Mackay}, {Mahony}, {McClure-Griffiths}, {McConigley}, {Mirtschin}, {Ng}, {Norris}, {Pearce}, {Phillips}, {Pilawa}, {Raja}, {Reynolds}, {Roberts}, {Roxby}, {Sadler}, {Shields}, {Schinckel}, {Serra}, {Shaw}, {Sweetnam}, {Troup}, {Tzioumis}, {Voronkov} and {Westmeier}}]{2021PASA...38....9H}
\bibinfo{author}{{Hotan}, A.W.}, \bibinfo{author}{{Bunton}, J.D.}, \bibinfo{author}{{Chippendale}, A.P.}, \bibinfo{author}{{Whiting}, M.}, \bibinfo{author}{{Tuthill}, J.}, \bibinfo{author}{{Moss}, V.A.}, \bibinfo{author}{{McConnell}, D.}, \bibinfo{author}{{Amy}, S.W.}, \bibinfo{author}{{Huynh}, M.T.}, \bibinfo{author}{{Allison}, J.R.}, \bibinfo{author}{{Anderson}, C.S.}, \bibinfo{author}{{Bannister}, K.W.}, \bibinfo{author}{{Bastholm}, E.}, \bibinfo{author}{{Beresford}, R.}, \bibinfo{author}{{Bock}, D.C.J.}, \bibinfo{author}{{Bolton}, R.}, \bibinfo{author}{{Chapman}, J.M.}, \bibinfo{author}{{Chow}, K.}, \bibinfo{author}{{Collier}, J.D.}, \bibinfo{author}{{Cooray}, F.R.}, \bibinfo{author}{{Cornwell}, T.J.}, \bibinfo{author}{{Diamond}, P.J.}, \bibinfo{author}{{Edwards}, P.G.}, \bibinfo{author}{{Feain}, I.J.}, \bibinfo{author}{{Franzen}, T.M.O.}, \bibinfo{author}{{George}, D.}, \bibinfo{author}{{Gupta}, N.}, \bibinfo{author}{{Hampson}, G.A.}, \bibinfo{author}{{Harvey-Smith}, L.}, \bibinfo{author}{{Hayman},
  D.B.}, \bibinfo{author}{{Heywood}, I.}, \bibinfo{author}{{Jacka}, C.}, \bibinfo{author}{{Jackson}, C.A.}, \bibinfo{author}{{Jackson}, S.}, \bibinfo{author}{{Jeganathan}, K.}, \bibinfo{author}{{Johnston}, S.}, \bibinfo{author}{{Kesteven}, M.}, \bibinfo{author}{{Kleiner}, D.}, \bibinfo{author}{{Koribalski}, B.S.}, \bibinfo{author}{{Lee-Waddell}, K.}, \bibinfo{author}{{Lenc}, E.}, \bibinfo{author}{{Lensson}, E.S.}, \bibinfo{author}{{Mackay}, S.}, \bibinfo{author}{{Mahony}, E.K.}, \bibinfo{author}{{McClure-Griffiths}, N.M.}, \bibinfo{author}{{McConigley}, R.}, \bibinfo{author}{{Mirtschin}, P.}, \bibinfo{author}{{Ng}, A.K.}, \bibinfo{author}{{Norris}, R.P.}, \bibinfo{author}{{Pearce}, S.E.}, \bibinfo{author}{{Phillips}, C.}, \bibinfo{author}{{Pilawa}, M.A.}, \bibinfo{author}{{Raja}, W.}, \bibinfo{author}{{Reynolds}, J.E.}, \bibinfo{author}{{Roberts}, P.}, \bibinfo{author}{{Roxby}, D.N.}, \bibinfo{author}{{Sadler}, E.M.}, \bibinfo{author}{{Shields}, M.}, \bibinfo{author}{{Schinckel}, A.E.T.},
  \bibinfo{author}{{Serra}, P.}, \bibinfo{author}{{Shaw}, R.D.}, \bibinfo{author}{{Sweetnam}, T.}, \bibinfo{author}{{Troup}, E.R.}, \bibinfo{author}{{Tzioumis}, A.}, \bibinfo{author}{{Voronkov}, M.A.}, \bibinfo{author}{{Westmeier}, T.}, \bibinfo{year}{2021}.
\newblock \bibinfo{title}{{Australian square kilometre array pathfinder: I. system description}}.
\newblock \bibinfo{journal}{PASA} \bibinfo{volume}{38}, \bibinfo{pages}{e009}.
\newblock \DOIprefix\doi{10.1017/pasa.2021.1}, \href{http://arxiv.org/abs/2102.01870}{{\tt arXiv:2102.01870}}.
\bibitem[{{Hurley-Walker} et~al.(2023){Hurley-Walker}, {Rea}, {McSweeney}, {Meyers}, {Lenc}, {Heywood}, {Hyman}, {Men}, {Clarke}, {Coti Zelati}, {Price}, {Horv{\'a}th}, {Galvin}, {Anderson}, {Bahramian}, {Barr}, {Bhat}, {Caleb}, {Dall'Ora}, {de Martino}, {Giacintucci}, {Morgan}, {Rajwade}, {Stappers} and {Williams}}]{2023Natur.619..487H}
\bibinfo{author}{{Hurley-Walker}, N.}, \bibinfo{author}{{Rea}, N.}, \bibinfo{author}{{McSweeney}, S.J.}, \bibinfo{author}{{Meyers}, B.W.}, \bibinfo{author}{{Lenc}, E.}, \bibinfo{author}{{Heywood}, I.}, \bibinfo{author}{{Hyman}, S.D.}, \bibinfo{author}{{Men}, Y.P.}, \bibinfo{author}{{Clarke}, T.E.}, \bibinfo{author}{{Coti Zelati}, F.}, \bibinfo{author}{{Price}, D.C.}, \bibinfo{author}{{Horv{\'a}th}, C.}, \bibinfo{author}{{Galvin}, T.J.}, \bibinfo{author}{{Anderson}, G.E.}, \bibinfo{author}{{Bahramian}, A.}, \bibinfo{author}{{Barr}, E.D.}, \bibinfo{author}{{Bhat}, N.D.R.}, \bibinfo{author}{{Caleb}, M.}, \bibinfo{author}{{Dall'Ora}, M.}, \bibinfo{author}{{de Martino}, D.}, \bibinfo{author}{{Giacintucci}, S.}, \bibinfo{author}{{Morgan}, J.S.}, \bibinfo{author}{{Rajwade}, K.M.}, \bibinfo{author}{{Stappers}, B.}, \bibinfo{author}{{Williams}, A.}, \bibinfo{year}{2023}.
\newblock \bibinfo{title}{{A long-period radio transient active for three decades}}.
\newblock \bibinfo{journal}{Nature} \bibinfo{volume}{619}, \bibinfo{pages}{487--490}.
\newblock \DOIprefix\doi{10.1038/s41586-023-06202-5}.
\bibitem[{{Huynh} et~al.(2020){Huynh}, {Dempsey}, {Whiting} and {Ophel}}]{2020ASPC..522..263H}
\bibinfo{author}{{Huynh}, M.}, \bibinfo{author}{{Dempsey}, J.}, \bibinfo{author}{{Whiting}, M.T.}, \bibinfo{author}{{Ophel}, M.}, \bibinfo{year}{2020}.
\newblock \bibinfo{title}{{The CSIRO ASKAP Science Data Archive}}, in: \bibinfo{editor}{{Ballester}, P.}, \bibinfo{editor}{{Ibsen}, J.}, \bibinfo{editor}{{Solar}, M.}, \bibinfo{editor}{{Shortridge}, K.} (Eds.), \bibinfo{booktitle}{Astronomical Data Analysis Software and Systems XXVII}, p. \bibinfo{pages}{263}.
\bibitem[{{Kemp} et~al.(2024){Kemp}, {Tingay}, {Midgley} and {Mitchell}}]{2024AJ....168..153K}
\bibinfo{author}{{Kemp}, I.}, \bibinfo{author}{{Tingay}, S.}, \bibinfo{author}{{Midgley}, S.}, \bibinfo{author}{{Mitchell}, D.}, \bibinfo{year}{2024}.
\newblock \bibinfo{title}{{An Image-based Blind Search for Fast Radio Bursts in 88 hr of Data from the EoR0 Field, with the Murchison Widefield Array}}.
\newblock \bibinfo{journal}{Astronomical Journal} \bibinfo{volume}{168}, \bibinfo{pages}{153}.
\newblock \DOIprefix\doi{10.3847/1538-3881/ad6f9c}, \href{http://arxiv.org/abs/2408.12200}{{\tt arXiv:2408.12200}}.
\bibitem[{{Kemp} et~al.(2025){Kemp}, {Pingel}, {Worth}, {Wake}, {Mitchell}, {Midgely}, {Tingay}, {Dempsey}, {D{\'e}nes}, {Dickey}, {Gibson}, {Jameson}, {Lynn}, {Ma}, {Marchal}, {McClure-Griffiths}, {Stanimirovi{\'c}} and {van Loon}}]{2025A&C....5100901K}
\bibinfo{author}{{Kemp}, I.P.}, \bibinfo{author}{{Pingel}, N.M.}, \bibinfo{author}{{Worth}, R.}, \bibinfo{author}{{Wake}, J.}, \bibinfo{author}{{Mitchell}, D.A.}, \bibinfo{author}{{Midgely}, S.D.}, \bibinfo{author}{{Tingay}, S.J.}, \bibinfo{author}{{Dempsey}, J.}, \bibinfo{author}{{D{\'e}nes}, H.}, \bibinfo{author}{{Dickey}, J.M.}, \bibinfo{author}{{Gibson}, S.J.}, \bibinfo{author}{{Jameson}, K.E.}, \bibinfo{author}{{Lynn}, C.}, \bibinfo{author}{{Ma}, Y.K.}, \bibinfo{author}{{Marchal}, A.}, \bibinfo{author}{{McClure-Griffiths}, N.M.}, \bibinfo{author}{{Stanimirovi{\'c}}, S.}, \bibinfo{author}{{van Loon}, J.T.}, \bibinfo{year}{2025}.
\newblock \bibinfo{title}{{Processing of GASKAP-Hi pilot survey data using a commercial supercomputer}}.
\newblock \bibinfo{journal}{Astronomy and Computing} \bibinfo{volume}{51}, \bibinfo{pages}{100901}.
\newblock \DOIprefix\doi{10.1016/j.ascom.2024.100901}, \href{http://arxiv.org/abs/2411.17118}{{\tt arXiv:2411.17118}}.
\bibitem[{Lee-Wadell et~al.(2022)Lee-Wadell, Pallott, German, Null, Shen, Aniruddha, Williamson and Holmes}]{AusSRC_DSP}
\bibinfo{author}{Lee-Wadell, K.}, \bibinfo{author}{Pallott, D.}, \bibinfo{author}{German, G.}, \bibinfo{author}{Null, D.}, \bibinfo{author}{Shen, A.}, \bibinfo{author}{Aniruddha, G.}, \bibinfo{author}{Williamson, A.}, \bibinfo{author}{Holmes, K.}, \bibinfo{year}{2022}.
\newblock \bibinfo{title}{Australian SKA Regional Centre Design Study Program: Final Report}.
\newblock \bibinfo{type}{Technical Report}. SKAO.
\newblock \URLprefix \url{https://aussrc.org/wp-content/uploads/2022/12/ DSPFinalReport\_29112022.pdf}.
\bibitem[{{Lonsdale} et~al.(2009){Lonsdale}, {Cappallo}, {Morales}, {Briggs}, {Benkevitch}, {Bowman}, {Bunton}, {Burns}, {Corey}, {Desouza}, {Doeleman}, {Derome}, {Deshpande}, {Gopala}, {Greenhill}, {Herne}, {Hewitt}, {Kamini}, {Kasper}, {Kincaid}, {Kocz}, {Kowald}, {Kratzenberg}, {Kumar}, {Lynch}, {Madhavi}, {Matejek}, {Mitchell}, {Morgan}, {Oberoi}, {Ord}, {Pathikulangara}, {Prabu}, {Rogers}, {Roshi}, {Salah}, {Sault}, {Shankar}, {Srivani}, {Stevens}, {Tingay}, {Vaccarella}, {Waterson}, {Wayth}, {Webster}, {Whitney}, {Williams} and {Williams}}]{2009IEEEP..97.1497L}
\bibinfo{author}{{Lonsdale}, C.J.}, \bibinfo{author}{{Cappallo}, R.J.}, \bibinfo{author}{{Morales}, M.F.}, \bibinfo{author}{{Briggs}, F.H.}, \bibinfo{author}{{Benkevitch}, L.}, \bibinfo{author}{{Bowman}, J.D.}, \bibinfo{author}{{Bunton}, J.D.}, \bibinfo{author}{{Burns}, S.}, \bibinfo{author}{{Corey}, B.E.}, \bibinfo{author}{{Desouza}, L.}, \bibinfo{author}{{Doeleman}, S.S.}, \bibinfo{author}{{Derome}, M.}, \bibinfo{author}{{Deshpande}, A.}, \bibinfo{author}{{Gopala}, M.R.}, \bibinfo{author}{{Greenhill}, L.J.}, \bibinfo{author}{{Herne}, D.E.}, \bibinfo{author}{{Hewitt}, J.N.}, \bibinfo{author}{{Kamini}, P.A.}, \bibinfo{author}{{Kasper}, J.C.}, \bibinfo{author}{{Kincaid}, B.B.}, \bibinfo{author}{{Kocz}, J.}, \bibinfo{author}{{Kowald}, E.}, \bibinfo{author}{{Kratzenberg}, E.}, \bibinfo{author}{{Kumar}, D.}, \bibinfo{author}{{Lynch}, M.J.}, \bibinfo{author}{{Madhavi}, S.}, \bibinfo{author}{{Matejek}, M.}, \bibinfo{author}{{Mitchell}, D.A.}, \bibinfo{author}{{Morgan}, E.}, \bibinfo{author}{{Oberoi}, D.},
  \bibinfo{author}{{Ord}, S.}, \bibinfo{author}{{Pathikulangara}, J.}, \bibinfo{author}{{Prabu}, T.}, \bibinfo{author}{{Rogers}, A.}, \bibinfo{author}{{Roshi}, A.}, \bibinfo{author}{{Salah}, J.E.}, \bibinfo{author}{{Sault}, R.J.}, \bibinfo{author}{{Shankar}, N.U.}, \bibinfo{author}{{Srivani}, K.S.}, \bibinfo{author}{{Stevens}, J.}, \bibinfo{author}{{Tingay}, S.}, \bibinfo{author}{{Vaccarella}, A.}, \bibinfo{author}{{Waterson}, M.}, \bibinfo{author}{{Wayth}, R.B.}, \bibinfo{author}{{Webster}, R.L.}, \bibinfo{author}{{Whitney}, A.R.}, \bibinfo{author}{{Williams}, A.}, \bibinfo{author}{{Williams}, C.}, \bibinfo{year}{2009}.
\newblock \bibinfo{title}{{The Murchison Widefield Array: Design Overview}}.
\newblock \bibinfo{journal}{IEEE Proceedings} \bibinfo{volume}{97}, \bibinfo{pages}{1497--1506}.
\newblock \DOIprefix\doi{10.1109/JPROC.2009.2017564}, \href{http://arxiv.org/abs/0903.1828}{{\tt arXiv:0903.1828}}.
\bibitem[{{McMullin} et~al.(2022){McMullin}, {Diamond}, {Caiazzo}, {Casson}, {Cheetham}, {Dewdney}, {Laing}, {Lewis}, {Schinckel}, {Stringhetti}, {Rees}, {van Es}, {Labate}, {Swart}, {Ball}, {Berry}, {Braun}, {Chrysostomou}, {Magnus} and {Pearce}}]{2022SPIE12182E..0QM}
\bibinfo{author}{{McMullin}, J.}, \bibinfo{author}{{Diamond}, P.}, \bibinfo{author}{{Caiazzo}, M.}, \bibinfo{author}{{Casson}, A.}, \bibinfo{author}{{Cheetham}, T.}, \bibinfo{author}{{Dewdney}, P.}, \bibinfo{author}{{Laing}, R.}, \bibinfo{author}{{Lewis}, B.}, \bibinfo{author}{{Schinckel}, A.}, \bibinfo{author}{{Stringhetti}, L.}, \bibinfo{author}{{Rees}, N.}, \bibinfo{author}{{van Es}, A.}, \bibinfo{author}{{Labate}, M.G.}, \bibinfo{author}{{Swart}, G.}, \bibinfo{author}{{Ball}, L.}, \bibinfo{author}{{Berry}, S.}, \bibinfo{author}{{Braun}, R.}, \bibinfo{author}{{Chrysostomou}, A.}, \bibinfo{author}{{Magnus}, L.}, \bibinfo{author}{{Pearce}, S.}, \bibinfo{year}{2022}.
\newblock \bibinfo{title}{{The Square Kilometre Array project update}}, in: \bibinfo{editor}{{Marshall}, H.K.}, \bibinfo{editor}{{Spyromilio}, J.}, \bibinfo{editor}{{Usuda}, T.} (Eds.), \bibinfo{booktitle}{Ground-based and Airborne Telescopes IX}, p. \bibinfo{pages}{121820Q}.
\newblock \DOIprefix\doi{10.1117/12.2642184}.
\bibitem[{Merriam and Tisdell(2015)}]{Merriam2015}
\bibinfo{author}{Merriam, S.B.}, \bibinfo{author}{Tisdell, E.J.}, \bibinfo{year}{2015}.
\newblock \bibinfo{title}{Qualitative Research: A Guide to Design and Implementation}.
\newblock \bibinfo{publisher}{Jossey-Bass}, \bibinfo{address}{San Francisco}.
\bibitem[{Mitchell(2024)}]{TechDay}
\bibinfo{author}{Mitchell, S.}, \bibinfo{year}{2024}.
\newblock \bibinfo{title}{{NVIDIA unveils AI SuperPOD, powered by high-efficiency Superchips}}.
\newblock \bibinfo{journal}{TechDay IT Brief} \URLprefix \url{https://itbrief.com.au/story/nvidia-unveils-ai-superpod-powered-by-high- efficiency-superchips}. \bibinfo{note}{[Accessed 21 Nov 2024]}.
\bibitem[{NHMRC(2018)}]{NHMRC}
\bibinfo{author}{NHMRC}, \bibinfo{year}{2018}.
\newblock \bibinfo{title}{National Statement on Ethical Conduct in Human Research 2007 (Updated 2018)}.
\newblock \bibinfo{publisher}{National Health and Medical Research Council}.
\newblock \URLprefix \url{http://www.nhmrc.gov.au/guidelines/publications/e72}.
\bibitem[{{Nikolic} and {SDP Consortium}(2014)}]{2014era..conf20201N}
\bibinfo{author}{{Nikolic}, B.}, \bibinfo{author}{{SDP Consortium}, S.}, \bibinfo{year}{2014}.
\newblock \bibinfo{title}{{Square Kilometre Array Science Data Processing}}, in: \bibinfo{booktitle}{Exascale Radio Astronomy}, p. \bibinfo{pages}{20201}.
\bibitem[{{Norris}(2011)}]{Norris2011}
\bibinfo{author}{{Norris}, R.P.}, \bibinfo{year}{2011}.
\newblock \bibinfo{title}{{Data Challenges for Next-generation Radio Telescopes}}, in: \bibinfo{booktitle}{Sixth {IEEE} International Conference on {eScience}}, pp. \bibinfo{pages}{21--24}.
\newblock \DOIprefix\doi{10.1109/eScienceW.2010.13}, \href{http://arxiv.org/abs/1101.1355}{{\tt arXiv:1101.1355}}.
\bibitem[{{Offringa} et~al.(2014){Offringa}, {McKinley}, {Hurley-Walker}, {Briggs}, {Wayth}, {Kaplan}, {Bell}, {Feng}, {Neben}, {Hughes}, {Rhee}, {Murphy}, {Bhat}, {Bernardi}, {Bowman}, {Cappallo}, {Corey}, {Deshpande}, {Emrich}, {Ewall-Wice}, {Gaensler}, {Goeke}, {Greenhill}, {Hazelton}, {Hindson}, {Johnston-Hollitt}, {Jacobs}, {Kasper}, {Kratzenberg}, {Lenc}, {Lonsdale}, {Lynch}, {McWhirter}, {Mitchell}, {Morales}, {Morgan}, {Kudryavtseva}, {Oberoi}, {Ord}, {Pindor}, {Procopio}, {Prabu}, {Riding}, {Roshi}, {Shankar}, {Srivani}, {Subrahmanyan}, {Tingay}, {Waterson}, {Webster}, {Whitney}, {Williams} and {Williams}}]{2014MNRAS.444..606O}
\bibinfo{author}{{Offringa}, A.R.}, \bibinfo{author}{{McKinley}, B.}, \bibinfo{author}{{Hurley-Walker}, N.}, \bibinfo{author}{{Briggs}, F.H.}, \bibinfo{author}{{Wayth}, R.B.}, \bibinfo{author}{{Kaplan}, D.L.}, \bibinfo{author}{{Bell}, M.E.}, \bibinfo{author}{{Feng}, L.}, \bibinfo{author}{{Neben}, A.R.}, \bibinfo{author}{{Hughes}, J.D.}, \bibinfo{author}{{Rhee}, J.}, \bibinfo{author}{{Murphy}, T.}, \bibinfo{author}{{Bhat}, N.D.R.}, \bibinfo{author}{{Bernardi}, G.}, \bibinfo{author}{{Bowman}, J.D.}, \bibinfo{author}{{Cappallo}, R.J.}, \bibinfo{author}{{Corey}, B.E.}, \bibinfo{author}{{Deshpande}, A.A.}, \bibinfo{author}{{Emrich}, D.}, \bibinfo{author}{{Ewall-Wice}, A.}, \bibinfo{author}{{Gaensler}, B.M.}, \bibinfo{author}{{Goeke}, R.}, \bibinfo{author}{{Greenhill}, L.J.}, \bibinfo{author}{{Hazelton}, B.J.}, \bibinfo{author}{{Hindson}, L.}, \bibinfo{author}{{Johnston-Hollitt}, M.}, \bibinfo{author}{{Jacobs}, D.C.}, \bibinfo{author}{{Kasper}, J.C.}, \bibinfo{author}{{Kratzenberg}, E.}, \bibinfo{author}{{Lenc},
  E.}, \bibinfo{author}{{Lonsdale}, C.J.}, \bibinfo{author}{{Lynch}, M.J.}, \bibinfo{author}{{McWhirter}, S.R.}, \bibinfo{author}{{Mitchell}, D.A.}, \bibinfo{author}{{Morales}, M.F.}, \bibinfo{author}{{Morgan}, E.}, \bibinfo{author}{{Kudryavtseva}, N.}, \bibinfo{author}{{Oberoi}, D.}, \bibinfo{author}{{Ord}, S.M.}, \bibinfo{author}{{Pindor}, B.}, \bibinfo{author}{{Procopio}, P.}, \bibinfo{author}{{Prabu}, T.}, \bibinfo{author}{{Riding}, J.}, \bibinfo{author}{{Roshi}, D.A.}, \bibinfo{author}{{Shankar}, N.U.}, \bibinfo{author}{{Srivani}, K.S.}, \bibinfo{author}{{Subrahmanyan}, R.}, \bibinfo{author}{{Tingay}, S.J.}, \bibinfo{author}{{Waterson}, M.}, \bibinfo{author}{{Webster}, R.L.}, \bibinfo{author}{{Whitney}, A.R.}, \bibinfo{author}{{Williams}, A.}, \bibinfo{author}{{Williams}, C.L.}, \bibinfo{year}{2014}.
\newblock \bibinfo{title}{{WSCLEAN: an implementation of a fast, generic wide-field imager for radio astronomy}}.
\newblock \bibinfo{journal}{MNRAS} \bibinfo{volume}{444}, \bibinfo{pages}{606--619}.
\newblock \DOIprefix\doi{10.1093/mnras/stu1368}, \href{http://arxiv.org/abs/1407.1943}{{\tt arXiv:1407.1943}}.
\bibitem[{{Pearson} et~al.(2025){Pearson}, {Ledford}, {Lutson} and {Van Noorden}}]{2025Pearson}
\bibinfo{author}{{Pearson}, H.}, \bibinfo{author}{{Ledford}, H.}, \bibinfo{author}{{Lutson}, M.}, \bibinfo{author}{{Van Noorden}, R.}, \bibinfo{year}{2025}.
\newblock \bibinfo{title}{{Exclusive: the most-cited papers of the twenty-first century}}.
\newblock \bibinfo{journal}{Nature} \bibinfo{volume}{640}, \bibinfo{pages}{588--593}.
\newblock \URLprefix \url{https://go.nature.com/4tsggtd}.
\bibitem[{Quartermaine and Kitaeff(2020)}]{AusSRC_Survey}
\bibinfo{author}{Quartermaine, L.}, \bibinfo{author}{Kitaeff, S.}, \bibinfo{year}{2020}.
\newblock \bibinfo{title}{AusSRC Radio Astronomy Data User Community Survey Report}.
\newblock \bibinfo{type}{Technial Report} \bibinfo{number}{AUSSRC-DSP-2020-0002}. SKAO.
\newblock \URLprefix \url{https://aussrc.org/ wp-content/uploads/2020/07/AusSRC-2019-Radio-Data-User- Survey-Report.pdf}.
\bibitem[{Saleh(2011)}]{Saleh2011}
\bibinfo{author}{Saleh, B.}, \bibinfo{year}{2011}.
\newblock \bibinfo{title}{Introduction to Subsurface Imaging}.
\newblock \bibinfo{publisher}{Cambridge University Press}, \bibinfo{address}{San Francisco}.
\newblock \DOIprefix\doi{10.1017/CBO9780511732577}.
\bibitem[{{Salgado} et~al.(2024){Salgado}, {Kitaeff}, {Bolton}, {Chrysostomou}, {Swinbank}, {Vilotte}, {Goliath}, {Gaudet}, {Barbosa}, {Yates}, {Ferrari}, {Beswick}, {Wadadekar}, {Bartolini} and {Rees}}]{2024ASPC..535..399S}
\bibinfo{author}{{Salgado}, J.}, \bibinfo{author}{{Kitaeff}, S.}, \bibinfo{author}{{Bolton}, R.}, \bibinfo{author}{{Chrysostomou}, A.}, \bibinfo{author}{{Swinbank}, J.}, \bibinfo{author}{{Vilotte}, J.}, \bibinfo{author}{{Goliath}, S.}, \bibinfo{author}{{Gaudet}, S.}, \bibinfo{author}{{Barbosa}, D.}, \bibinfo{author}{{Yates}, J.}, \bibinfo{author}{{Ferrari}, C.}, \bibinfo{author}{{Beswick}, R.}, \bibinfo{author}{{Wadadekar}, Y.}, \bibinfo{author}{{Bartolini}, M.}, \bibinfo{author}{{Rees}, N.}, \bibinfo{year}{2024}.
\newblock \bibinfo{title}{{Designing the SKA Regional Centre Network}}, in: \bibinfo{editor}{{Hugo}, B.V.}, \bibinfo{editor}{{Van Rooyen}, R.}, \bibinfo{editor}{{Smirnov}, O.M.} (Eds.), \bibinfo{booktitle}{Astromical Data Analysis Software and Systems XXXI}, p. \bibinfo{pages}{399}.
\bibitem[{{Sault} et~al.(1995){Sault}, {Teuben} and {Wright}}]{1995ASPC...77..433S}
\bibinfo{author}{{Sault}, R.J.}, \bibinfo{author}{{Teuben}, P.J.}, \bibinfo{author}{{Wright}, M.C.H.}, \bibinfo{year}{1995}.
\newblock \bibinfo{title}{{A Retrospective View of MIRIAD}}, in: \bibinfo{editor}{{Shaw}, R.A.}, \bibinfo{editor}{{Payne}, H.E.}, \bibinfo{editor}{{Hayes}, J.J.E.} (Eds.), \bibinfo{booktitle}{Astronomical Data Analysis Software and Systems IV}, p. \bibinfo{pages}{433}.
\newblock \DOIprefix\doi{10.48550/arXiv.astro-ph/0612759}, \href{http://arxiv.org/abs/astro-ph/0612759}{{\tt arXiv:astro-ph/0612759}}.
\bibitem[{{Schinckel} et~al.(2012){Schinckel}, {Bunton}, {Cornwell}, {Feain} and {Hay}}]{2012SPIE.8444E..2AS}
\bibinfo{author}{{Schinckel}, A.E.}, \bibinfo{author}{{Bunton}, J.D.}, \bibinfo{author}{{Cornwell}, T.J.}, \bibinfo{author}{{Feain}, I.}, \bibinfo{author}{{Hay}, S.G.}, \bibinfo{year}{2012}.
\newblock \bibinfo{title}{{The Australian SKA Pathfinder}}, in: \bibinfo{editor}{{Stepp}, L.M.}, \bibinfo{editor}{{Gilmozzi}, R.}, \bibinfo{editor}{{Hall}, H.J.} (Eds.), \bibinfo{booktitle}{Ground-based and Airborne Telescopes IV}, p. \bibinfo{pages}{84442A}.
\newblock \DOIprefix\doi{10.1117/12.926959}.
\bibitem[{{Servers.com}(2025)}]{Servers2025}
\bibinfo{author}{{Servers.com}}, \bibinfo{year}{2025}.
\newblock \bibinfo{title}{Bare metal vs cloud servers:the right infrastructure matters}.
\newblock \URLprefix \url{https://www.servers.com/news/blog/bare-metal-vs-cloud-servers-what-s-the-difference}. \bibinfo{note}{[Accessed: 31 May 2025]}.
\bibitem[{SKAO(2024a)}]{SKAO}
\bibinfo{author}{SKAO}, \bibinfo{year}{2024}a.
\newblock \bibinfo{title}{Handling a deluge of big data}.
\newblock \URLprefix \url{https://www.skao.int/en/explore/big-data}. \bibinfo{note}{[Accessed 24 Mar 2025]}.
\bibitem[{SKAO(2024b)}]{SKAOSRC}
\bibinfo{author}{SKAO}, \bibinfo{year}{2024}b.
\newblock \bibinfo{title}{Ska regional centres}.
\newblock \URLprefix \url{https://www.skao.int/en/science-users/119/ska-regional-centres}. \bibinfo{note}{[Accessed 6 Jun 2025]}.
\bibitem[{Sleap(2025)}]{MWAX}
\bibinfo{author}{Sleap, G.}, \bibinfo{year}{2025}.
\newblock \bibinfo{title}{{MWAX Modes}}.
\newblock \URLprefix \url{https://mwatelescope. atlassian.net/wiki/spaces/MP/pages/24973003/MWAX+Modes}. \bibinfo{note}{[Accessed 24 Mar 2025]}.
\bibitem[{{The National Research Infrastructure Advisory Group}(2024)}]{NDRIS}
\bibinfo{author}{{The National Research Infrastructure Advisory Group}}, \bibinfo{year}{2024}.
\newblock \bibinfo{title}{National Digital Research Infrastructure Strategy}.
\newblock \bibinfo{publisher}{Australian Government}, \bibinfo{address}{Canberra, Australia}.
\newblock \URLprefix \url{https://www.education.gov.au/national-research-infrastructure/resources/national- digital-research-infrastructure-strategy}.
\bibitem[{{Tingay} et~al.(2013){Tingay}, {Goeke}, {Bowman}, {Emrich}, {Ord}, {Mitchell}, {Morales}, {Booler}, {Crosse}, {Wayth}, {Lonsdale}, {Tremblay}, {Pallot}, {Colegate}, {Wicenec}, {Kudryavtseva}, {Arcus}, {Barnes}, {Bernardi}, {Briggs}, {Burns}, {Bunton}, {Cappallo}, {Corey}, {Deshpande}, {Desouza}, {Gaensler}, {Greenhill}, {Hall}, {Hazelton}, {Herne}, {Hewitt}, {Johnston-Hollitt}, {Kaplan}, {Kasper}, {Kincaid}, {Koenig}, {Kratzenberg}, {Lynch}, {Mckinley}, {Mcwhirter}, {Morgan}, {Oberoi}, {Pathikulangara}, {Prabu}, {Remillard}, {Rogers}, {Roshi}, {Salah}, {Sault}, {Udaya-Shankar}, {Schlagenhaufer}, {Srivani}, {Stevens}, {Subrahmanyan}, {Waterson}, {Webster}, {Whitney}, {Williams}, {Williams} and {Wyithe}}]{2013PASA...30....7T}
\bibinfo{author}{{Tingay}, S.J.}, \bibinfo{author}{{Goeke}, R.}, \bibinfo{author}{{Bowman}, J.D.}, \bibinfo{author}{{Emrich}, D.}, \bibinfo{author}{{Ord}, S.M.}, \bibinfo{author}{{Mitchell}, D.A.}, \bibinfo{author}{{Morales}, M.F.}, \bibinfo{author}{{Booler}, T.}, \bibinfo{author}{{Crosse}, B.}, \bibinfo{author}{{Wayth}, R.B.}, \bibinfo{author}{{Lonsdale}, C.J.}, \bibinfo{author}{{Tremblay}, S.}, \bibinfo{author}{{Pallot}, D.}, \bibinfo{author}{{Colegate}, T.}, \bibinfo{author}{{Wicenec}, A.}, \bibinfo{author}{{Kudryavtseva}, N.}, \bibinfo{author}{{Arcus}, W.}, \bibinfo{author}{{Barnes}, D.}, \bibinfo{author}{{Bernardi}, G.}, \bibinfo{author}{{Briggs}, F.}, \bibinfo{author}{{Burns}, S.}, \bibinfo{author}{{Bunton}, J.D.}, \bibinfo{author}{{Cappallo}, R.J.}, \bibinfo{author}{{Corey}, B.E.}, \bibinfo{author}{{Deshpande}, A.}, \bibinfo{author}{{Desouza}, L.}, \bibinfo{author}{{Gaensler}, B.M.}, \bibinfo{author}{{Greenhill}, L.J.}, \bibinfo{author}{{Hall}, P.J.}, \bibinfo{author}{{Hazelton}, B.J.},
  \bibinfo{author}{{Herne}, D.}, \bibinfo{author}{{Hewitt}, J.N.}, \bibinfo{author}{{Johnston-Hollitt}, M.}, \bibinfo{author}{{Kaplan}, D.L.}, \bibinfo{author}{{Kasper}, J.C.}, \bibinfo{author}{{Kincaid}, B.B.}, \bibinfo{author}{{Koenig}, R.}, \bibinfo{author}{{Kratzenberg}, E.}, \bibinfo{author}{{Lynch}, M.J.}, \bibinfo{author}{{Mckinley}, B.}, \bibinfo{author}{{Mcwhirter}, S.R.}, \bibinfo{author}{{Morgan}, E.}, \bibinfo{author}{{Oberoi}, D.}, \bibinfo{author}{{Pathikulangara}, J.}, \bibinfo{author}{{Prabu}, T.}, \bibinfo{author}{{Remillard}, R.A.}, \bibinfo{author}{{Rogers}, A.E.E.}, \bibinfo{author}{{Roshi}, A.}, \bibinfo{author}{{Salah}, J.E.}, \bibinfo{author}{{Sault}, R.J.}, \bibinfo{author}{{Udaya-Shankar}, N.}, \bibinfo{author}{{Schlagenhaufer}, F.}, \bibinfo{author}{{Srivani}, K.S.}, \bibinfo{author}{{Stevens}, J.}, \bibinfo{author}{{Subrahmanyan}, R.}, \bibinfo{author}{{Waterson}, M.}, \bibinfo{author}{{Webster}, R.L.}, \bibinfo{author}{{Whitney}, A.R.}, \bibinfo{author}{{Williams}, A.},
  \bibinfo{author}{{Williams}, C.L.}, \bibinfo{author}{{Wyithe}, J.S.B.}, \bibinfo{year}{2013}.
\newblock \bibinfo{title}{{The Murchison Widefield Array: The Square Kilometre Array Precursor at Low Radio Frequencies}}.
\newblock \bibinfo{journal}{PASA} \bibinfo{volume}{30}, \bibinfo{pages}{e007}.
\newblock \DOIprefix\doi{10.1017/pasa.2012.007}, \href{http://arxiv.org/abs/1206.6945}{{\tt arXiv:1206.6945}}.
\bibitem[{Top500.org(2024)}]{TOP500}
\bibinfo{author}{Top500.org}, \bibinfo{year}{2024}.
\newblock \bibinfo{title}{Top 500 the list.}
\newblock \URLprefix \url{https://www.top500.org/}. \bibinfo{note}{[Accessed 24 Mar 2025]}.
\bibitem[{{Trader}(2021)}]{Trader2021}
\bibinfo{author}{{Trader}, T.}, \bibinfo{year}{2021}.
\newblock \bibinfo{title}{{HPCWire: AMD Launches Epyc ‘Milan’ with 19 SKUs for HPC, Enterprise and Hyperscale}}.
\newblock \URLprefix \url{https://www.hpcwire.com/2021/03/15/amd-launches-epyc-milan-with-19-skus-for-hpc-enterprise-and-hyperscale/}. \bibinfo{note}{[Accessed: 9 Mar 2025]}.
\bibitem[{{Wayth} et~al.(2018){Wayth}, {Tingay}, {Trott}, {Emrich}, {Johnston-Hollitt}, {McKinley}, {Gaensler}, {Beardsley}, {Booler}, {Crosse}, {Franzen}, {Horsley}, {Kaplan}, {Kenney}, {Morales}, {Pallot}, {Sleap}, {Steele}, {Walker}, {Williams}, {Wu}, {Cairns}, {Filipovic}, {Johnston}, {Murphy}, {Quinn}, {Staveley-Smith}, {Webster} and {Wyithe}}]{2018PASA...35...33W}
\bibinfo{author}{{Wayth}, R.B.}, \bibinfo{author}{{Tingay}, S.J.}, \bibinfo{author}{{Trott}, C.M.}, \bibinfo{author}{{Emrich}, D.}, \bibinfo{author}{{Johnston-Hollitt}, M.}, \bibinfo{author}{{McKinley}, B.}, \bibinfo{author}{{Gaensler}, B.M.}, \bibinfo{author}{{Beardsley}, A.P.}, \bibinfo{author}{{Booler}, T.}, \bibinfo{author}{{Crosse}, B.}, \bibinfo{author}{{Franzen}, T.M.O.}, \bibinfo{author}{{Horsley}, L.}, \bibinfo{author}{{Kaplan}, D.L.}, \bibinfo{author}{{Kenney}, D.}, \bibinfo{author}{{Morales}, M.F.}, \bibinfo{author}{{Pallot}, D.}, \bibinfo{author}{{Sleap}, G.}, \bibinfo{author}{{Steele}, K.}, \bibinfo{author}{{Walker}, M.}, \bibinfo{author}{{Williams}, A.}, \bibinfo{author}{{Wu}, C.}, \bibinfo{author}{{Cairns}, I.H.}, \bibinfo{author}{{Filipovic}, M.D.}, \bibinfo{author}{{Johnston}, S.}, \bibinfo{author}{{Murphy}, T.}, \bibinfo{author}{{Quinn}, P.}, \bibinfo{author}{{Staveley-Smith}, L.}, \bibinfo{author}{{Webster}, R.}, \bibinfo{author}{{Wyithe}, J.S.B.}, \bibinfo{year}{2018}.
\newblock \bibinfo{title}{{The Phase II Murchison Widefield Array: Design overview}}.
\newblock \bibinfo{journal}{PASA} \bibinfo{volume}{35}, \bibinfo{pages}{e033}.
\newblock \DOIprefix\doi{10.1017/pasa.2018.37}, \href{http://arxiv.org/abs/1809.06466}{{\tt arXiv:1809.06466}}.

\end{thebibliography}

\appendix

\section{Interview Guide}
\label{Survey_Questions}

\textbf{Section 1: Initial Triage to determine which section of the questionnaire applies}\\
1.	Please confirm that you’ve been involved in research or production data processing which make use of commercial High Performance Computing (HPC) ?\\
2.	Were you part of the research team, or part of the HPC providers organisation?\\
3.	Were you actually carrying out the research, or involved in a management capacity?\\
4.	Was the project in astronomy, or a difference scientific area?\\
\\
\textbf{Section 2: Questions for Researchers}\\
1.	Please briefly describe the projects you have been involved in that have used commercial HPC
<prompt if not mentioned>\\
a.	Who was involved?\\
b.	Which commercial provider?\\
c.	What was your personal role in the project?\\
2.	What were the key software tools used?\\
$<$prompt if not volunteered$>$\\
a.	Inversion/imaging/calibration\\
b.	Other analysis\\
c.	Workflow management\\
d.	Visualisation\\
3.	Why did the researchers choose to use a commercial provider?\\
4.	How did you set up the computing environment?
$<$prompt if not volunteered$>$
a.	Who did the software installs, astronomers or software engineers from either academia or the computing provider?\\
5.	Please describe some of the problems, if any, that arose.\\
a.	Did you solve them / How did you solve them?\\
b.	What level of assistance did you get from the provider?\\
$<$prompt if not volunteered$>$\\
c.	Data transport/ingest issues?\\
d.	Software issues?\\
e.	Compute availability/technical support?\\
6.	Please describe some of the benefits / advantages that you found, if any.\\
$<$prompt if not volunteered$>$\\
a.	Software support?\\
b.	Reliability of compute and/or storage?\\
c.	Tech support?\\
d.	Timeliness\\
7.	What do you think about the issue of cost?\\
8.	To use commercial providers effectively, what skills are needed on the astronomers’ side?\\
9.	To effectively work with researchers, what skills are needed on the providers’ side?\\
10.	What do you think is the future role of commercial supercomputing?\\
\\
\textbf{Section 3: Questions for HPC providers}\\
1.	Please briefly describe one of the projects you have been involved in that have used commercial HPC\\
$<$prompt if not mentioned$>$\\
a.	Who was involved?\\
b.	Which commercial provider?\\
c.	What was your personal role in the project?\\
2.	What were the key software tools used?\\
$<$prompt if not volunteered$>$\\
a.	Inversion/imaging/calibration\\
b.	Other analysis\\
c.	Workflow management\\
d.	Visualisation\\
3.	Why did the researchers choose to use a commercial provider?\\
4.	How did you set up the computing environment including the specialist tools?\\
$<$prompt if not volunteered$>$\\
a.	Who did the software installs, astronomers or software engineers from either academia or the computing provider?\\
5.	Please describe some of the problems, if any, that arose.\\
a.	Did you solve them / How did you solve them?\\
b.	What level of assistance did you get from the provider?\\
$<$prompt if not volunteered$>$\\
c.	Data transport/ingest issues?\\
d.	Software issues?\\
e.	Compute availability/technical support?\\
6.	Please describe some of the benefits / advantages that you found, if any.\\
$<$prompt if not volunteered$>$\\
a.	Software support?\\
b.	Reliability of compute and/or storage?\\
c.	Tech support?\\
d.	Timeliness\\
7.	What do you think about the issue of cost?\\
8.	To effectively use commercial providers, what skills are needed on the astronomers’ side?\\
9.	To effectively work with researchers, what skills are needed on the providers’ side?\\
10.	What do you think is the future role of commercial supercomputing?\\
\\
\textbf{Section 4: Questions for Research Managers}\\
1.	Where do you think the future lies with HPC in astronomy and other scientific research – in-house HPC, partnership, or will HPC become a commodity?\\
2.	What are the main skills or services that commercial providers need to offer in order to effectively engage with researchers?\\
3.	What do you perceive as the main differences between in-house or government-funded HPC services and commercial providers?\\
4.	What are some of the difficulties in working with the commercial providers?\\
5.	Are there any legal issues that arise?\\
6.	What do you think are the main benefits that commercial providers can offer researchers?\\
$<$prompt for comment if not mentioned$>$\\
a.	Volume of work?\\
b.	Timeliness?\\
c.	Specialist capability?\\
d.	Cost?\\

\section{Codes}
\label{Codes}
Communication\\
Cost, budgetting\\
Data Management\\
Funding\\
HPC specialists\\
Legal issues\\
Limited Storage\\
Node selection\\
Overflow capacity\\
Parallelism\\
Platform selection\\
Project Management\\
Reliability\\
Security management\\
Skills\\
Software\\
Time allocation

\end{document}